\documentclass[10pt, prb, aps, twocolumn, showpacs, citeautoscript, floatfix, reprint, amsmath, amssymb, notitlepage, superscriptaddress]{revtex4-2}

\usepackage{graphicx}
\usepackage{stmaryrd}
\usepackage{rotating}
\usepackage{amsmath}
\usepackage{amsfonts}
\usepackage{amssymb}
\usepackage{bbm}
\usepackage{wasysym}
\usepackage[countmax]{subfloat}
\usepackage{dcolumn} 
\usepackage{bm} 
\usepackage{color}
\usepackage[colorlinks=true, allcolors=blue]{hyperref}
\usepackage[compatibility=false]{caption}
\usepackage{subcaption}

\usepackage{float}
\usepackage{latexsym}
\usepackage{booktabs}

\begin{document}

\title{Quantum geometric bounds and uncertainty relations for velocity variance \\ 
and spread of Wannier function}






\author{Felipe P. Abreu}

\affiliation{Department of Physics, PUC-Rio, 22451-900 Rio de Janeiro, Brazil}

\author{Luis F. C\'{a}rdenas-Castillo}

\affiliation{Department of Physics, PUC-Rio, 22451-900 Rio de Janeiro, Brazil}

\author{David Porlles}

\affiliation{Department of Physics, PUC-Rio, 22451-900 Rio de Janeiro, Brazil}

\author{Wei Chen}

\affiliation{Department of Physics, PUC-Rio, 22451-900 Rio de Janeiro, Brazil}

\date{\rm\today}

\begin{abstract}

By utilizing the derivation of quantum Cram\'{e}r-Rao bound and an argument put forward by Efimov and Dodonov, we derive several uncertainty relations for mixed states that constrain the diagonal product, trace, and each diagonal element of the covariance matrix of multiple observables. By choosing the observables to be the symmetric logarithmic derivative, the uncertainty relations imply various geometric bounds on quantum Fisher information matrix caused by mean Uhlmann curvature, which in pure states become geometric bounds on quantum metric caused by Berry curvature. Our formalism reveals optical sum rules for velocity variance and spread of valence band Wannier functions, and how they are bounded by the Berry curvature and commutator of velocity operators. Canonical ensembles of spins, topological materials, common semiconductors, and transition metal dichalcogenides are used to verify these geometric bounds and uncertainty relations.

\end{abstract}

\maketitle

\section{Introduction}

The quantum metric derived from the fidelity between neighboring Bloch states has risen to prominence recently in condensed matter physics, partially owing to its relation with dielectric and optical properties in semiconductors and insulators\cite{Ozawa18,Ahn22,Komissarov24,Chen25_optical_marker}. Besides these remarkable features of quantum metric itself, since quantum metric and Berry curvature are real and imaginary parts of the quantum geometric tensor\cite{Provost80,Berry84}, the interplay between the two has also been an intriguing issue that triggered a fair amount of investigation. Along this direction, several constraints on quantum metric caused by Berry curvature have been discovered, of which we collectively call quantum geometric bounds. For instance, in 2D momentum spaces, the determinant of quantum metric is bounded from below by the square of Berry curvature\cite{Roy14,Peotta15,Ozawa21}, rendering a constraint on the volume form of the Euclidean manifold. This bound is saturated in 2D topological materials\cite{Ma13,Ma14,Yang15,Piechon16,Palumbo18,Panahiyan20_fidelity_susceptibility,Ma20,Salerno20,Mera22}, giving rise to a metric-curvature correspondence that may help to detect the topological order\cite{vonGersdorff21_metric_curvature}. Recently, another bound on each diagonal element of quantum metric caused by Berry curvature has also been discovered\cite{Chen26_QCRB}, pointing to an even stronger constraint imposed by Berry curvature.

Despite these intriguing relations between quantum metric and Berry curvature uncovered recently, it should be reminded that many of these geometric aspects are actually not new, but already developed decades ago within the context of quantum metrology that concerns systems described by density matrices instead of just quantum states. For instance, quantum metric is actually the pure-state limit of the quantum Fisher information matrix (QFIM) derived from the fidelity between density matrices at neighboring points in the parameter space\cite{Helstrom67,Hovelo11,Braunstein94,Braunstein96,Paris09,Liu20}. Similarly, Berry curvature can be viewed as the pure-state and Abelian limit of the mean Uhlmann curvature (MUC) that characterizes the geometric phase accumulated by bringing a density matrix along an infinitesimal loop\cite{Uhlmann86,Uhlmann91,Carollo18}. The QFIM and MUC are real and imaginary parts of the quantum Fisher tensor\cite{Ercolessi12,Carollo18}, and both can be derived from a generating function\cite{Hetenyi23,Hetenyi26,Abreu26_generating_fn}, rendering a unified picture for the mixed-state quantum metrology. Adopting this metrological point of view, it is then important to clarify whether the aforementioned geometric bounds have some deeper metrological interpretation, and whether certain metrological properties of QFIM and MUC can be applied to the quantum geometry of Bloch states.

In this paper, we elaborate that a concept in quantum metrology is particularly relevant to derive geometric bounds in a general setting, namely the uncertainty relations\cite{Robertson29,Schrodinger30,Hovelo11}. For both pure and mixed states, multi-observable uncertainty relations are certain constraints on the diagonal product, trace, or each diagonal element of the covariant matrix of multiple observables caused by the commutators of the observables, and many such relations have been derived by different means. In present work, we employ two different methods, one from the derivation of quantum Cram\'{e}r-Rao bound (QCRB)\cite{Helstrom67,Braunstein94,Braunstein96,Liu20} and the other from an argument put forward by Efimov and Dodonov\cite{Efimov76,Dodonov18}, to derive a set of uncertainty relations applicable to both pure and mixed states. We then observe that because QFIM and MUC are the covariance and commutators of the symmetric logarithmic derivative (SLD) stemmed from the density matrix, these uncertainty relations imply a set of geometric bounds on QFIM caused by MUC, suggesting strong constraints that MUC imposes on the problem of quantum parameter estimation. In addition, because QFIM and MUC reduce to quantum metric and Berry curvature in pure states, as mentioned above, our result also unveils a set of geometric bounds on the diagonal product, trace, and each diagonal element of quantum metric caused by Berry curvature, pointing to a highly constrained Euclidean manifold for systems that have nonzero Berry curvature.

The significance of these geometric bounds and uncertainty relations is elaborated for a variety of condensed matter systems. For semiconductors and insulators treated as canonical ensembles, the mixed-state uncertainty relations are shown to constrain observables at finite temperature, as demonstrated by the covariance of velocity operators in lattice models of transition metal dichalcogenides and common semiconductors\cite{Liu13,Vogl83,Graf95}. The bound on QFIM caused by MUC is verified in the canonical ensemble of spin $1/2$, which also reveals the constraint on the magnetic susceptibility as a QFIM in the magnetic field space. For the geometric bounds on quantum metric caused by Berry curvature in pure states, we provide an analytical verification in theoretical models of topological materials, and show that they have profound experimental consequences on optical sum rules of absorbance in 2D and dielectric function in 3D\cite{CardenasCastillo24_spread_Wannier}. Consequently, the spread of valence band Wannier function is found to be bounded in semiconductors that have nonzero Berry curvature\cite{Marzari97,Souza08,Marzari12}.

\section{Generator scenario for mixed-state quantum geometry}

\subsection{Uncertainty relations based on QCRB \label{sec:uncertainty_mixed_states}}


We first elaborate an uncertainty relation between a set of Hermitian operators, proved in a way similar to the derivation of QCRB by utilizing the generators of isospectral density matrices. Consider a density matrix $\rho({\bf k})$ defined in a $D$-dimensional parameter space ${\bf k}=(k^{1},k^{2}...k^{D})$, and a set of Hermitian operators ${\cal A}_{\mu}=\left\{{\cal A}_{1},{\cal A}_{2}...\right\}$ that may or may not depend on the parameters. For the sake of our discussion, we define the complex covariance $K_{\mu\nu}$, covariance $C_{\mu\nu}$, and commutator $B_{\mu\nu}$ of the Hermitian operators ${\cal A}_{\mu}$ by
\begin{eqnarray}
&&K_{\mu\nu}=\langle\Delta {\cal A}_{\mu}\Delta {\cal A}_{\nu}\rangle=C_{\mu\nu}-\frac{i}{2}B_{\mu\nu},
\nonumber \\
&&C_{\mu\nu}=\frac{1}{2}\langle\left\{\Delta {\cal A}_{\mu},\Delta {\cal A}_{\nu}\right\}\rangle,
\nonumber \\
&&B_{\mu\nu}=i\langle[{\cal A}_{\mu},{\cal A}_{\nu}]\rangle=
i\langle[\Delta{\cal A}_{\mu},\Delta{\cal A}_{\nu}]\rangle,\;\;\;\;\;
\label{Kmunu_Dmunu_Bmunu_A}
\end{eqnarray}
where $\Delta {\cal A}_\mu \equiv {\cal A}_\mu - \langle {\cal A}_\mu \rangle$ and $\langle{\hat O}\rangle\equiv{\rm Tr}\rho\,{\hat O}$.

Our proof starts by considering the Cauchy-Schwarz inequality 
\begin{eqnarray}
&&{\rm Tr} (X^\dagger X) {\rm Tr} (Y^\dagger Y) 
\nonumber \\
&&\geq \frac{1}{4} \left| {\rm Tr} (X^\dagger Y + X Y^\dagger) \right|^2+\frac{1}{4} \left| {\rm Tr} (X^\dagger Y - X Y^\dagger) \right|^2,\;\;\;\;\;
\label{CS_inequality_general_XY}
\end{eqnarray}
which leads to the following inequality 
\begin{eqnarray}
{\rm Tr} (X^\dagger X) {\rm Tr} (Y^\dagger Y) \geq \frac{1}{4} \left| {\rm Tr} (X^\dagger Y - X Y^\dagger) \right|^2.
\label{consequence_of_CS_inequality_minus}
\end{eqnarray}
Note that in the derivation of QCRB, one takes the $\frac{1}{4} \left| {\rm Tr} (X^\dagger Y + X Y^\dagger) \right|^2$ term in Eq.~(\ref{CS_inequality_general_XY}), but we take the term with a negative sign $\frac{1}{4} \left| {\rm Tr} (X^\dagger Y - X Y^\dagger) \right|^2$ in the calculation below. Now consider the following $X$ and $Y$ for the Cauchy-Schwarz inequality (repeating Greek indices are summed)
\begin{eqnarray}
X = f^\mu \Delta{\cal A}_\mu \sqrt{\rho},\;\;\;
Y = g^{\nu}\Delta{\cal A}_{\nu} \sqrt{\rho}
\label{XY_definition}
\end{eqnarray}
where $f^{\mu}$ and $g^{\nu}$ are arbitrary real numbers at this stage. Then 
\begin{eqnarray}
&&{\rm Tr} (X^\dagger X) 
= f^{\mu}C_{\mu\nu}f^{\nu}={\bf f}^{\rm T}\cdot{\bf C}\cdot{\bf f},
\nonumber \\
&&{\rm Tr} (Y^\dagger Y) = g^{\mu}C_{\mu\nu}g^{\nu}
={\bf g}^{T}\cdot{\bf C}\cdot{\bf g},
\end{eqnarray}
and right hand side of Eq.~(\ref{consequence_of_CS_inequality_minus}) becomes 
\begin{eqnarray}
&&\frac{1}{2} {\rm Tr} (X^\dagger Y - X Y^\dagger) = f^{\mu}\frac{1}{2}\langle[{\cal A}_{\mu},{\cal A}_{\nu}]\rangle g^{\nu}
\nonumber \\
&&=-\frac{i}{2}{\bf f}^{T}\cdot{\bf B}\cdot{\bf g},
\end{eqnarray}
Thus our inequality in Eq.~(\ref{consequence_of_CS_inequality_minus}) becomes
\begin{eqnarray}
{\bf f}^{\rm T}\cdot{\bf C}\cdot{\bf f}\times{\bf g}^{T}\cdot{\bf C}\cdot{\bf g}\geq\frac{1}{4}\left({\bf f}^{\rm T}\cdot{\bf B}\cdot{\bf g}\right)^2.
\label{inequality_fFf}
\end{eqnarray}
We now choose the real ${\bf f}$ vector to be
\begin{eqnarray}
{\bf f}={\bf C}^{-1}\cdot{\bf B}\cdot{\bf g},
\label{f_vector}
\end{eqnarray}
where ${\bf C}^{-1}$ is the inverse of the covariance matrix of the generators, leading to 
\begin{eqnarray}
{\bf g}^{T}\cdot{\bf C}\cdot{\bf g}\geq \frac{1}{4}{\bf g}^{T}\cdot\left({\bf B}^{T}\cdot{\bf C}^{-1}\cdot {\bf B}\right)\cdot{\bf g},\;\;\;
\label{gCg_gFg_inequality}
\end{eqnarray}
meaning that the matrix ${\bf C}-\frac{1}{4}{\bf B}^{T}\cdot{\bf C}^{-1}\cdot {\bf B}$ is positive semidefinitive. Together with the fact that ${\bf C}^{-1}$ is also positive semidefinitive, and ${\bf B}^{T}=-{\bf B}$ according to Eq.~(\ref{Kmunu_Dmunu_Bmunu_A}), we are lead to conclude that 
\begin{eqnarray}
{\bf C}\geq -\frac{1}{4}{\bf B}\cdot{\bf C}^{-1}\cdot {\bf B}\geq 0,
\label{QCRB_operator_mixed_state_generator}
\end{eqnarray}
or equivalently the diagonal elements of this inequality satisfy
\begin{eqnarray}
C_{\alpha\alpha}\geq-\frac{1}{4}B_{\alpha\mu}C^{\mu\nu}B_{\nu\alpha}.
\label{selfbound_geometric_interpretation_generator}
\end{eqnarray}
Note that the inverse matrix $C^{\mu\nu}$ is the adjugate matrix divided by the determinant
\begin{eqnarray}
{\bf C}^{-1}=\frac{1}{\det\bf{C}}\,{\rm adj}({\bf C}),
\label{ginverse_adjgdetg}
\end{eqnarray}
so at the end Eq.~(\ref{selfbound_geometric_interpretation_generator}) can always be written as a geometric bound on the lower indexed $C_{\mu\nu}$. Further invoking Eq.~(\ref{Kmunu_Dmunu_Bmunu_A}) leads to the following uncertainty relations for mixed states
\begin{eqnarray}
\langle\Delta{\cal A}_{\alpha}^{2}\rangle\geq\frac{1}{4}\langle[{\cal A}_{\alpha},{\cal A}_{\mu}]\rangle
\left(\frac{1}{2}\langle\left\{\Delta{\cal A},\Delta{\cal A}\right\}\rangle\right)^{\mu\nu}
\langle[{\cal A}_{\nu},{\cal A}_{\alpha}]\rangle.
\nonumber \\
\label{selfbound_uncertainty_interpretation_generator}
\end{eqnarray}
where $\left(\frac{1}{2}\langle\left\{\Delta{\cal A},\Delta{\cal A}\right\}\rangle\right)^{\mu\nu}$ stands for the inverse of the covariance matrix ${\bf C}$. We call this result the single-observable uncertainty relations because the left hand side consists of only one observable, and it is the mixed-state generalization of the same relations recently discovered for pure states\cite{Chen26_QCRB}.


To be specific, in 2D parameter space $\mu=\left\{x,y\right\}$, Eq.~(\ref{selfbound_geometric_interpretation_generator}) renders the inequality 
\begin{eqnarray}
&&C_{xx}\geq\frac{1}{4}B_{xy}^{2}C^{yy}=\frac{B_{xy}^{2}C_{xx}}{4(C_{xx}C_{yy}-C_{xy}^{2})}\geq 0,
\nonumber \\
&&C_{yy}\geq\frac{1}{4}B_{xy}^{2}C^{xx}=\frac{B_{xy}^{2}C_{yy}}{4(C_{xx}C_{yy}-C_{xy}^{2})}\geq 0.
\label{Dmumu_bound_2D}
\end{eqnarray}
After canceling out $C_{\mu\mu}$ and using Eq.~(\ref{Kmunu_Dmunu_Bmunu_A}), either one of the two inequalities in Eq.~(\ref{Dmumu_bound_2D}) yields
\begin{eqnarray}
&&\langle\Delta{\cal A}_{x}^{2}\rangle\langle\Delta{\cal A}_{y}^{2}\rangle\geq
\frac{1}{4}|\langle\left\{\Delta{\cal A}_{x},\Delta{\cal A}_{y}\right\}\rangle|^{2}
+\frac{1}{4}|\langle\left[{\cal A}_{x},{\cal A}_{y}\right]\rangle|^{2},
\nonumber \\
\label{Robertson_Schrodinger_Heisenberg_mixed_states}
\end{eqnarray}
rendering the well-known mixed-state Robertson-Schr\"{o}dinger uncertainty relation between two operators ${\cal A}_{x}$ and ${\cal A}_{y}$\cite{Robertson29,Schrodinger30,Hovelo11}. In other words, for two observables, Eq.~(\ref{selfbound_uncertainty_interpretation_generator}) is identical to the Robertson-Schr\"{o}dinger uncertainty relation. Following the same reasoning, the index-explicit expressions for Eqs.~(\ref{selfbound_geometric_interpretation_generator}) and (\ref{selfbound_uncertainty_interpretation_generator}) in 3D parameter spaces are given in Appendix \ref{apx:index_explicity_inequalities_3D}.

\subsection{Uncertainty relations based on Efimov-Dodonov argument \label{sec:Efimov_Dodonov_argument}}

We proceed to derive several more uncertainty relations for mixed states based on the argument put forward by Efimov and Dodonov\cite{Efimov76,Dodonov18}. Consider the following operator
\begin{eqnarray}
{\hat {\cal O}}=\sum_{\mu=x,y,z}\alpha_{\mu}\Delta{\cal A}_{\mu}\otimes\sigma_{\mu},
\label{O_operator_A}
\end{eqnarray}
where $\alpha_{\mu}$ are arbitrary real numbers, and $\sigma_{\mu}$ are the Pauli matrices. The operator is defined on an enlarged Hilbert space $|\Psi_{i}\rangle=|i\rangle\otimes|\chi\rangle$, with $|i\rangle$ as the eigenstates of density matrices that $\Delta{\cal A}_{\mu}$ acts upon, and $|\chi\rangle$ is the auxiliary spinor of the Pauli matrices $\sigma_{\mu}$. We then introduce a density matrix $\overline{\rho}$ in this enlarged Hilbert space from the density matrix $\rho=\sum_{i}\lambda_{i}|i\rangle\langle i|$ of the quantum system at hand by 
\begin{eqnarray}
\overline{\rho}=\sum_{i}\lambda_{i}|\Psi_{i}\rangle\langle\Psi_{i}|=
\sum_{i}\lambda_{i}|i\rangle\otimes|\chi\rangle\langle\chi|\otimes\langle i|.
\end{eqnarray}
The average of ${\hat {\cal O}}^{\dag}{\hat {\cal O}}$ with respect to $\overline{\rho}$ must be positive
\begin{eqnarray}
\langle {\hat {\cal O}}^{\dag}{\hat {\cal O}}\rangle
={\rm Tr}\overline{\rho}\,{\hat {\cal O}}^{\dag}{\hat {\cal O}}
=\sum_{i}\lambda_{i}\langle\Psi_{i}|{\hat {\cal O}}^{\dag}{\hat {\cal O}}|\Psi_{i}\rangle\geq 0,
\end{eqnarray}
since $\langle\Psi_{i}|{\hat {\cal O}}^{\dag}{\hat {\cal O}}|\Psi_{i}\rangle\geq 0$ and $\lambda_{i}\geq 0$. On the other hand, this average is also equal to
\begin{eqnarray}
&&\langle {\hat {\cal O}}^{\dag}{\hat {\cal O}}\rangle=\langle\chi|\left\{\left[\alpha_{x}^{2}C_{xx}+\alpha_{y}^{2}C_{yy}+\alpha_{z}^{2}C_{zz}\right]\sigma_{0}\right.
\nonumber \\
&&\left.+\alpha_{x}\alpha_{y}B_{xy}\sigma_{z}+\alpha_{y}\alpha_{z}B_{yz}\sigma_{x}+\alpha_{z}\alpha_{x}B_{zx}\sigma_{y}\right\}|\chi\rangle
\nonumber \\
&&=\langle\chi|f|\chi\rangle\geq 0,
\end{eqnarray}
implying that the matrix $f$ in the parentheses $\left\{...\right\}$ must be positive semidefinitive. As a result, one has $\det f\geq 0$, leading to 
\begin{eqnarray}
&&\alpha_{x}^{2}C_{xx}+\alpha_{y}^{2}C_{yy}+\alpha_{z}^{2}C_{zz}
\nonumber \\
&&\geq\left[(\alpha_{x}\alpha_{y}B_{xy})^{2}+(\alpha_{y}\alpha_{z}B_{yz})^{2}
+(\alpha_{z}\alpha_{x}B_{zx})^{2}\right]^{1/2}.
\nonumber \\
\label{Efimov_Dodonov_formula}
\end{eqnarray}
Several uncertainty relations can be derived from different choices of $\alpha_{\mu}$\cite{Dodonov18}, as we shall see below.

Firstly, choosing $\alpha_{x}=\alpha_{y}=\alpha_{z}=1$ results in a three-observable uncertainty relation that bounds the sum of the variances by the commutators
\begin{eqnarray}
&&\langle\Delta{\cal A}_{x}^{2}\rangle+\langle\Delta{\cal A}_{y}^{2}\rangle+\langle\Delta{\cal A}_{z}^{2}\rangle
\nonumber \\
&&\geq\left[\left|\langle[{\cal A}_{x},{\cal A}_{y}]\rangle\right|^{2}+\left|\langle[{\cal A}_{y},{\cal A}_{z}]\rangle\right|^{2}+\left|\langle[{\cal A}_{z},{\cal A}_{x}]\rangle\right|^{2}\right]^{1/2}.
\nonumber \\
\label{sum_bound_3D_commutator}
\end{eqnarray}
Secondly, choosing $\alpha_{x}=\alpha_{y}=1$ and $\alpha_{z}=0$ results in a two-observable uncertainty relation that also bounds the sum of the variances by the only commutator
\begin{eqnarray}
\langle\Delta{\cal A}_{x}^{2}\rangle+\langle\Delta{\cal A}_{y}^{2}\rangle\geq\left|\langle[{\cal A}_{x},{\cal A}_{y}]\rangle\right|.
\label{sum_bound_2D_commutator}
\end{eqnarray}
Besides these bounds for the sum, Eq.~(\ref{Efimov_Dodonov_formula}) also obtains a three-observable uncertainty relation for the product of the variances by choosing $\alpha_{x}^{2}=C_{yy}C_{zz}$, $\alpha_{y}^{2}=C_{zz}C_{xx}$, $\alpha_{z}^{2}=C_{xx}C_{yy}$, which takes the form
\begin{eqnarray}
&&\langle\Delta{\cal A}_{x}^{2}\rangle\langle\Delta{\cal A}_{y}^{2}\rangle\langle\Delta{\cal A}_{z}^{2}\rangle
\geq\frac{1}{9}\left\{\langle\Delta{\cal A}_{x}^{2}\rangle\left|\langle[{\cal A}_{y},{\cal A}_{z}]\rangle\right|^{2}\right.
\nonumber \\
&&\left.+\langle\Delta{\cal A}_{y}^{2}\rangle\left|\langle[{\cal A}_{z},{\cal A}_{x}]\rangle\right|^{2}+\langle\Delta{\cal A}_{z}^{2}\rangle\left|\langle[{\cal A}_{x},{\cal A}_{y}]\rangle\right|^{2}\right\}.\;\;\;
\label{product_bound_3D_commutator}
\end{eqnarray}
Similarly, choosing $\alpha_{x}^{2}=C_{yy}$, $\alpha_{y}^{2}=C_{xx}$, and $\alpha_{z}^{2}=0$ yields an uncertainty relation for two observables
\begin{eqnarray}
&&\langle\Delta{\cal A}_{x}^{2}\rangle\langle\Delta{\cal A}_{y}^{2}\rangle
\geq\frac{1}{4}\left|\langle[{\cal A}_{x},{\cal A}_{y}]\rangle\right|^{2}.
\label{product_bound_2D_Heisenberg}
\end{eqnarray}
that gives the mixed-state Heisenberg uncertainty relation, which is a weaker bound than Eq.~(\ref{Robertson_Schrodinger_Heisenberg_mixed_states}). The pure state version of these uncertainty relations are those originally proposed by Dodonov\cite{Dodonov18}, and here we demonstrate that they remain valid for mixed states.


\subsection{Geometric bounds for mixed states \label{sec:geometric_bounds_mixed_states}}

In this section, we elaborate that any uncertainty relations consisting of covariances $C_{\mu\nu}$ and commutators $B_{\mu\nu}$ always imply geometric bounds on QFIM and MUC. To see this, we start by the SLD operator $L_{\mu}$ defined from the derivative of the density matrix
\begin{eqnarray}
    \partial_{\mu}\rho=\frac{1}{2} \{ \rho, L_{\mu}\},
\label{SLD_definition}
\end{eqnarray}
from which the elements of the quantum Fisher tensor $T_{\mu\nu}$, QFIM $F_{\mu\nu}$, and MUC ${\cal U}_{\mu\nu}$ are introduced as the expectation values of products, covariances, and commutators of SLD, respectively
\begin{eqnarray}
&&T_{\mu\nu}=\frac{1}{4}\langle L_{\mu}L_{\nu}\rangle=\frac{1}{4}\langle \Delta L_{\mu}\Delta L_{\nu}\rangle=\frac{1}{4}F_{\mu\nu}-\frac{i}{2}{\cal U}_{\mu\nu},
\nonumber \\
&&F_{\mu\nu}=\frac{1}{2}\langle\left\{L_{\mu},L_{\nu}\right\}\rangle
=\frac{1}{2}\langle\left\{\Delta L_{\mu},\Delta L_{\nu}\right\}\rangle, 
\nonumber \\
&&{\cal U}_{\mu\nu}=\frac{i}{4}\langle[L_{\mu},L_{\nu}]\rangle=\frac{i}{4}\langle[\Delta L_{\mu},\Delta L_{\nu}]\rangle.
\label{QFIM_MUC_SLD_definition}
\end{eqnarray}
where we have used $\langle L_{\mu}\rangle=0$. By comparing Eqs.~(\ref{Kmunu_Dmunu_Bmunu_A}) and (\ref{QFIM_MUC_SLD_definition}), it is then immediately obvious that any uncertainty relation for mixed states becomes a geometric bound on QFIM and MUC if we simply choose the observable under question to be the SLD ${\cal A}_{\mu}=L_{\mu}$, i.e.,
\begin{eqnarray}
&&{\rm Uncertainty}\left.\left(\frac{1}{2}\langle\left\{\Delta {\cal A}_{\mu},\Delta {\cal A}_{\nu}\right\}\rangle,i\langle[{\cal A}_{\mu},{\cal A}_{\nu}]\rangle\right)\right|_{{\cal A}_{\mu}=L_{\mu}}
\nonumber \\
&&\Rightarrow{\rm Bound}\left(F_{\mu\nu},4\,{\cal U}_{\mu\nu}\right).
\label{bound_uncertainty_correspondence_mixed_state}
\end{eqnarray}
For example, Eqs.~(\ref{QCRB_operator_mixed_state_generator}) and (\ref{selfbound_geometric_interpretation_generator}) imply that 
\begin{eqnarray}
&&{\bf F}\geq -4\,{\cal U}\cdot{\bf F}^{-1}\cdot {\cal U}\geq 0,
\nonumber \\
&&F_{\alpha\alpha}\geq-4\,{\cal U}_{\alpha\mu}F^{\mu\nu}{\cal U}_{\nu\alpha},
\label{Fmumu_UFU_bound_general}
\end{eqnarray}
that bounds the elements of QFIM and MUC in parameter space of arbitrary dimensions. Applying the same calculation in Eq.~(\ref{Dmumu_bound_2D}) to 2D parameter spaces $\mu=\left\{x,y\right\}$, Eq.~(\ref{Fmumu_UFU_bound_general}) leads to 
\begin{eqnarray}
F_{xx} F_{yy} \geq F_{xy}^2 + 4\,{\cal U}_{xy}^2.
\label{FxxFyy_Uxy_bound}
\end{eqnarray}
The index-explicit expressions of Eq.~(\ref{Fmumu_UFU_bound_general}) in 3D parameter spaces $\mu=\left\{x,y,z\right\}$ are given in Appendix \ref{apx:index_explicity_inequalities_3D}.

For $2\times 2$ density matrices, one may verify the geometric bounds analytically by means of Bloch representation
\begin{align}
    \rho = \frac{1}{2}(\mathbb{I}_2 + \bm{r}\cdot \bm{\sigma}),
\label{rho_Bloch_2by2}
\end{align}
where $\bm{r}$ is the Bloch vector ($|\bm{r}|^2 \leq 1$) and $\bm{\sigma}$ are the Pauli matrices, leading to the QFIM and MUC\cite{Watanabe10,Watanabe11,Liu20,Abreu26_generating_fn}
\begin{align}
    F_{\mu\nu} &= \partial_{\mu}{\bm r}\cdot\partial_{\nu}{\bm r}
    +\frac{({\bm r}\cdot\partial_{\mu}{\bm r})({\bm r}\cdot\partial_{\nu}{\bm r})}{1-|{\bm r}|^{2}} 
    \\
    {\cal U}_{\mu\nu} &= -\frac{1}{2}{\bm r}\cdot\partial_{\mu}{\bm r}\times\partial_{\nu}{\bm r}.
\label{QFIM_MUC_2by2_Bloch}
\end{align}
As a result, the 3D bound given in Eq.\eqref{selfbound_3C_index_explicit} has the generic expression
\begin{align}
    F_{\mu\mu} = |\partial_{\mu}{\bm r}|^2
    +\frac{({\bm r}\cdot\partial_{\mu}{\bm r})^2}{1-|{\bm r}|^{2}} \geq | {\bm r} \times \partial_\mu {\bm r}|^2 \geq 0,
\end{align}
for each $\mu = \{x,y,z\}$, which is valid because
\begin{align}
    \label{Bound_mumu}
    |\partial_{\mu}{\bm r}|^2(1 - |\bm{r}|^2) + ({\bm r}\cdot\partial_{\mu}{\bm r})^2\left(
    \frac{2-|{\bm r}|^{2}}{1-|{\bm r}|^{2}}\right) \geq 0.
\end{align}
Furthermore, the 2D bound in Eq.~(\ref{FxxFyy_Uxy_bound}) requires that 
\begin{align}
    &F_{xx} F_{yy} - F_{xy}^2 - 4\,{\cal U}_{xy}^2 \\ 
    &= \frac{|\partial_x \bm{r} \times \partial_y \bm{r} |^2 - (2-r^2)[\bm{r} \cdot(\partial_x \bm{r} \times \partial_y \bm{r})]^2}{1-|\bm{r}|^2} \geq 0. \nonumber
\end{align}
Given that $[\bm{r} \cdot(\partial_x \bm{r} \times \partial_y \bm{r})]^2 = |\bm{r}|^2 |\partial_x \bm{r} \times \partial_y \bm{r}|^2\cos^{2}{\theta}$, where $\theta$ is the angle between the vectors $\bm{r}$ and $\partial_x \bm{r} \times \partial_y \bm{r}$. The lowest value for the bound occurs for $\cos^{2} \theta=1$, at which the numerator becomes $|\partial_x \bm{r} \times \partial_y \bm{r} |^2(1-|\bm{r}|^2)^2 \geq 0$, and hence the $2D$ bound is also satisfied.


Following the same reasoning, choosing the generators in Eq.~(\ref{O_operator_A}) to be the SLD ${\cal A}_{\mu}=L_{\mu}$ leads to 
\begin{eqnarray}
&&\alpha_{x}^{2}F_{xx}+\alpha_{y}^{2}F_{yy}+\alpha_{z}^{2}F_{zz}
\nonumber \\
&&\geq\left[(4\alpha_{x}\alpha_{y}{\cal U}_{xy})^{2}+(4\alpha_{y}\alpha_{z}{\cal U}_{yz})^{2}
+(4\alpha_{z}\alpha_{x}{\cal U}_{zx})^{2}\right]^{1/2}.
\nonumber \\
\end{eqnarray}
that is analogous to Eq.~(\ref{Efimov_Dodonov_formula}). Choosing $\alpha_{x}=\alpha_{y}=\alpha_{z}=1$ results in a bound on the trace of QFIM caused by MUC on 3D manifolds
\begin{eqnarray}
\label{Bound_sum_Fxx_Fyy_Fzz}
F_{xx}+F_{yy}+F_{zz}\geq 4\left[{\cal U}_{xy}^{2}+{\cal U}_{yz}^{2}+{\cal U}_{zx}^{2}\right]^{1/2}.
\end{eqnarray}
Similarly, choosing $\alpha_{x}=\alpha_{y}=1$ and $\alpha_{z}=0$ gives a bound on the trace of QFIM on 2D manifolds
\begin{eqnarray}
\label{Bound_sum_Fxx_Fyy}
F_{xx}+F_{yy}\geq 4|{\cal U}_{xy}|.
\end{eqnarray}
Furthermore, by $\alpha_{x}^{2}=F_{yy}F_{zz}$, $\alpha_{y}^{2}=F_{zz}F_{xx}$, $\alpha_{z}^{2}=F_{xx}F_{yy}$, we see that the diagonal product of QFIM on 3D manifolds is bounded by 
\begin{eqnarray}
\label{Bound_product_Fxx_Fyy_Fzz}
F_{xx}F_{yy}F_{zz}\geq \frac{16}{9}\left(F_{xx}{\cal U}_{yz}^{2}+F_{yy}{\cal U}_{zx}^{2}+F_{zz}{\cal U}_{xy}^{2}\right).
\end{eqnarray}
Analogously, the choice $\alpha_{x}^{2}=F_{yy}$, $\alpha_{y}^{2}=F_{xx}$, and $\alpha_{z}^{2}=0$ leads to the bound on the diagonal product of QFIM on 2D manifolds
\begin{eqnarray}
\label{Bound_product_Fxx_Fyy}
F_{xx}F_{yy}\geq 4{\cal U}_{xy}^{2},
\end{eqnarray}
which is a weaker bound than Eq.~(\ref{FxxFyy_Uxy_bound}). These geometric bounds on QFIM imposed by MUC are valid for any density matrices defined in any 2D and 3D parameter spaces.

\subsection{Application: Canonical ensemble of spin \texorpdfstring{$1/2$}{1/2}}

To elaborate the applications in mixed states, we consider the canonical ensemble of spin $1/2$ in a magnetic field described by the $2\times 2$ Hamiltonian $H=-{\bm B}\cdot{\boldsymbol\sigma}$ with the magnetic field ${\bm B}=(B_{x},B_{y},B_{z})$ pointing in an arbitrary direction, and $\beta=1/k_{B}T$ is determined by the temperature $T$. The density matrix $\rho=e^{-\beta H}/{\rm Tr}e^{-\beta H}$ in the form of Eq.~(\ref{rho_Bloch_2by2}) is parametrized by ${\bm r}=({\bm B}/B)\tanh\beta B$, where $B=(B_{x}^{2}+B_{y}^{2}+B_{z}^{2})^{1/2}$ is the magnitude of the magnetic field. We treat the magnetic field $(B_{x},B_{y},B_{z})$ as a $3D$ Euclidean manifold, so $\partial_{\mu}\equiv\partial/\partial B_{\mu}$. Equation (\ref{QFIM_MUC_2by2_Bloch}) then gives
\begin{align}
&F_{\mu\nu} =  \frac{\tanh^2(\beta B)}{B^2}\left( \delta_{\mu\nu} -\frac{B_\mu B_\nu}{B^2}\right) +   \beta^2 \text{sech}^2(\beta B) \frac{B_\mu B_\nu}{B^2}
\nonumber \\
&{\cal U}_{\mu\nu} =-\frac{\varepsilon_{\mu\nu\sigma}B_{\sigma}\tanh^{3}\beta B}{2B^{3}},
\label{QFIM_MUC_spin_half}
\end{align}
where $\varepsilon_{\mu\nu\sigma}$ is the Levi-Civita symbol. Note that the magnetic susceptibility tensor is given by
\begin{eqnarray}
&&\chi_{\mu\nu} =\frac{\partial \langle \sigma_\mu\rangle}{\partial B_\nu}
\nonumber \\
&&=  \frac{\tanh(\beta B)}{B}\left( \delta_{\mu\nu} -\frac{B_\mu B_\nu}{B^2}\right) +   \beta \text{sech}^2(\beta B) \frac{B_\mu B_\nu}{B^2}. 
\nonumber \\
\end{eqnarray}
Comparing the QFIM in Eq.~(\ref{QFIM_MUC_spin_half}), we conclude that
\begin{align}
    \chi_{\mu\nu} = \sqrt{F_T}\left( \delta_{\mu\nu} -\frac{B_\mu B_\nu}{B^2}\right) +   \frac{F_L}{\beta}\frac{B_\mu B_\nu}{B^2},
\end{align}
where $F_T = \tanh^2{(\beta B)}/B^2$ and $F_L = \beta^2{\rm sech}^2(\beta B)$ are the transverse and longitudinal eigenvalues of the QFIM, and $ \delta_{\mu\nu} - B_\mu B_\nu/B^2$ and $B_\mu B_\nu/B^2$ are the transverse and longitudinal projectors of the magnetic field, i.e., 
\begin{eqnarray}
F_{\mu\nu} =  F_T\left( \delta_{\mu\nu} -\frac{B_\mu B_\nu}{B^2}\right) +  F_L\frac{B_\mu B_\nu}{B^2}.
\end{eqnarray}
If the magnetic field is only along one direction $\bm{B} = B \hat{\mu}$, the relation between the QFIM and the magnetic susceptibility simplifies to $\chi_{\mu\mu} = F_{\mu\mu}/\beta$. In this sense, the magnetic susceptibility is bounded by the measurement incompatibility between independent components of the magnetic field, from which such incompatibility arises due to the quantum nature of the system\cite{Carollo18,Ragy16}.



\begin{figure}[ht]
\begin{center}
\includegraphics[clip=true,width=0.99\columnwidth]{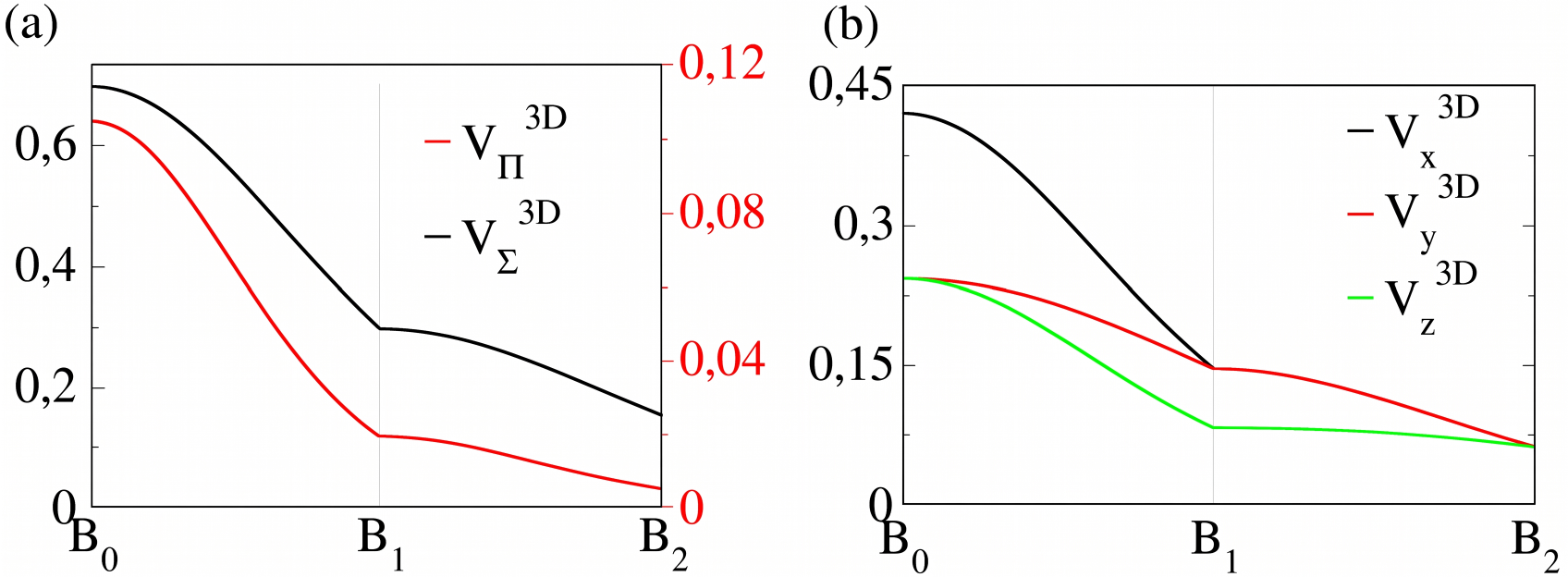}
\caption{Evolution of the bounds on the QFIM components along the path $B_{0}-B_{1}-B_{2}$ in the magnetic field space, where $B_0=(1,0,0)$, $B_1=(1,1,0)$, and $B_2=(1,1,1)$, setting $\beta=1/k_{B}T=1$ as the energy unit. (a) Summation bound $V_{\Sigma}^{3D}$ for Eq.~(\ref{Bound_sum_Fxx_Fyy_Fzz}) (black curve, left scale), and the product bound $V_{\Pi}^{3D}$ for Eq.~(\ref{Bound_product_Fxx_Fyy_Fzz}) (red curve, right scale). (b) Individual bounds $V_{\mu}^{3D}$ in Eq.~(\ref{selfbound_3C_index_explicit}) for $\mu = \{x,y,z\}=\{{\rm black,red,green}\}$.} 
\label{fig:Bound_QFIM_spin_half}
\end{center}
\end{figure}

To corroborate the geometric bounds on QFIM caused by MUC, in Fig.~\ref{fig:Bound_QFIM_spin_half} we examine these bounds along the line $B_{0}-B_{1}-B_{2}$ on the 3D manifold of magnetic field, where $B_0=(1,0,0)$, $B_1=(1,1,0)$, and $B_2=(1,1,1)$. In Fig.~\ref{fig:Bound_QFIM_spin_half} (a), we plot the $V_{\Sigma}^{3D}$ defined as the left hand side minus the right hand side of the summation bound in Eq.~(\ref{Bound_sum_Fxx_Fyy_Fzz}), and similarly $V_{\Pi}^{3D}$ defined as the left hand side minus the right hand side of the product bound in Eq.~(\ref{Bound_product_Fxx_Fyy_Fzz}). In Fig.~\ref{fig:Bound_QFIM_spin_half} (b), we inspect the bound on individual elements by plotting the left hand side minus the right hands side of each inequality in  Eq.~(\ref{selfbound_3C_index_explicit}) denoted by $V_{\mu}^{3D}$. All these quantities are positive everywhere on the manifold, indicating the validity of these bounds.

\subsection{Pure state limit of geometric bounds \label{sec:pure_state_limit_geometric_bounds}}

For pure states, the density matrix $\rho=|\psi\rangle\langle\psi|$ contains only one quantum state $|\psi({\bf k})\rangle$, and the SLD in Eq.~(\ref{SLD_definition}) is solved by $L_{\mu}=2|\partial_{\mu}\psi\rangle\langle\psi|+2|\psi\rangle\langle\partial_{\mu}\psi|$. As a result, quantum Fisher tensor reduces to the quantum geometric tensor $T_{\mu\nu}$, a quarter of QFIM $F_{\mu\nu}/4$ reduces to the quantum metric $g_{\mu\nu}$, and MUC ${\cal U}_{\mu\nu}$ reduces to the Berry curvature $\Omega_{\mu\nu}$, i.e.,
\begin{eqnarray}
T_{\mu\nu}\rightarrow T_{\mu\nu}&=&\langle\partial_{\mu}\psi|\partial_{\nu}\psi\rangle
-\langle\partial_{\mu}\psi|\psi\rangle\langle\psi|\partial_{\nu}\psi\rangle
\nonumber \\
&=&g_{\mu\nu}-\frac{i}{2}\Omega_{\mu\nu}
\nonumber \\
\frac{F_{\mu\nu}}{4}\rightarrow g_{\mu\nu}&=&\frac{1}{2}\langle\partial_{\mu}\psi|\partial_{\nu}\psi\rangle
+\frac{1}{2}\langle\partial_{\nu}\psi|\partial_{\mu}\psi\rangle
\nonumber \\
&-&\langle\partial_{\mu}\psi|\psi\rangle\langle\psi|\partial_{\nu}\psi\rangle
\nonumber \\
{\cal U}_{\mu\nu}\rightarrow\Omega_{\mu\nu}&=&i\langle\partial_{\mu}\psi|\partial_{\nu}\psi\rangle-
i\langle\partial_{\nu}\psi|\partial_{\mu}\psi\rangle.
\label{gmunu_Omegamunu_formula}
\end{eqnarray}
As a result, the geometric bounds discovered in Sec.~\ref{sec:geometric_bounds_mixed_states} recover the bounds on quantum metric and Berry curvature discovered in our recent work\cite{Chen26_QCRB}. Similarly, the geometric bounds for 2D and 3D manifolds listed in Sec.~\ref{sec:Efimov_Dodonov_argument} become
\begin{eqnarray}
&&g_{xx}+g_{yy}+g_{zz}\geq \left[\Omega_{xy}^{2}+\Omega_{yz}^{2}+\Omega_{zx}^{2}\right]^{1/2}.
\nonumber \\
&&g_{xx}+g_{yy}\geq |\Omega_{xy}|.
\nonumber \\
&&g_{xx}g_{yy}g_{zz}\geq \frac{1}{9}\left(g_{xx}\Omega_{yz}^{2}+g_{yy}\Omega_{zx}^{2}+g_{zz}\Omega_{xy}^{2}\right).
\nonumber \\
&&g_{xx}g_{yy}\geq \frac{1}{4}\Omega_{xy}^{2}.
\label{bounds_gmumu_sum_prod}
\end{eqnarray}
Note that the last constraint is a weaker bound on $g_{xx}g_{yy}$ than the well-known bound $g_{xx}g_{yy}\geq g_{xy}^{2}+\Omega_{xy}^{2}/4$ that has been derived by different means\cite{Roy14,Peotta15,Ozawa21}. Finally, the uncertainty relations in Sec.~\ref{sec:uncertainty_mixed_states} are still valid in pure states and recover those discovered recently\cite{Chen26_QCRB}, and the uncertainty relations in Sec.~\ref{sec:Efimov_Dodonov_argument} recover those discovered by Dodonov\cite{Dodonov18}, where the meaning of expectation value is changed to $\langle{\hat{\cal O}}\rangle=\langle\psi|{\hat{\cal O}}|\psi\rangle$.

\section{Applications to semiconductors\label{sec:applications_semiconductors}}

\subsection{Bound on spread of Wannier functions and optical sum rules}

We proceed to elaborate the geometric bounds as constraints on the spread of Wannier function and optical sum rules. For semiconductors, insulators, or superconductors at zero temperature that have $N_{-}$ valence or quasihole bands, a quantum metric can be constructed from the fully antisymmetric filled band Bloch state
\begin{eqnarray}
|\psi({\bf k})\rangle\equiv \frac{1}{\sqrt{ N_-!}}\epsilon^{a_1\cdots a_{N_-}}|n_{a_1}\rangle|n_{a_2}\rangle\cdots|n_{a_{N_{-}}}\rangle.
\label{eq:fermi}
\end{eqnarray}
by treating the crystalline momentum ${\bf k}=(k^{x},k^{y}...k^{D})$ as the tuning parameter. This results in a quantum metric and Berry curvature of the form\cite{Matsuura10,vonGersdorff21_metric_curvature}
\begin{eqnarray}
&&g_{\mu\nu}=\frac{1}{2}\sum_{nm}\langle\partial_{\mu}n|m\rangle\langle m|\partial_{\nu}n\rangle+(\mu\leftrightarrow \nu),
\nonumber \\
&&\Omega_{\mu\nu}=i\sum_{nm}\langle\partial_{\mu}n|m\rangle\langle m|\partial_{\nu}n\rangle-(\mu\leftrightarrow \nu),
\label{gmunu_Omegamunu_nm_expression}
\end{eqnarray}
where $|n\rangle$ and $|m\rangle$ denote the valence (or quasihole) and conduction (or quasiparticle) band states, respectively. The spread of Wannier function of valence or quasihole bands is defined by\cite{Marzari97,Marzari12,Porlles23_singlet_SC_metric}
\begin{eqnarray}
&&\Omega=\sum_{n}\left[\langle{\bf 0}n|r^{2}|{\bf 0}n\rangle-\langle{\bf 0}n|{\bf r}|{\bf 0}n\rangle^{2}\right]=\Omega_{I}+\tilde{\Omega},
\label{Omega_original}
\end{eqnarray}
which can be separated to the gauge-invariant $\Omega_{I}$ and gauge-dependent $\tilde{\Omega}$, where $|{\bf R}n\rangle$ is the Wannier state for filled band $n$ centering at homecell ${\bf R}$, and ${\bf r}$ is the position operator. Defining the fidelity number as the momentum integration of the quantum metric\cite{deSousa23_fidelity_marker}
\begin{eqnarray}
&&{\cal G}_{\mu\nu}=\int\frac{d^{D}{\bf k}}{(2\pi)^{D}}g_{\mu\nu}({\bf k}),
\label{Gmunu_definition}
\end{eqnarray} 
it is known that the gauge-invariant part of the spread is given by the trace of the fidelity number\cite{Marzari97,Marzari12,Souza08}, and hence are bounded in 2D and 3D by
\begin{eqnarray}
&&\Omega_{I}^{2D}
=A_{\rm cell}{\rm Tr}\,{\cal G}_{\mu\nu}\geq A_{\rm cell}\int\frac{d^{2}{\bf k}}{(2\pi)^{2}}|\Omega_{xy}|,
\nonumber \\
&&\Omega_{I}^{3D}
=\frac{V_{\rm cell}}{\hbar}{\rm Tr}\,{\cal G}_{\mu\nu}
\nonumber \\
&&\;\;\;\;\;\;\;\;\geq\frac{V_{\rm cell}}{\hbar}\int\frac{d^{3}{\bf k}}{(2\pi)^{3}}[\Omega_{xy}^{2}+\Omega_{yz}^{2}+\Omega_{zx}^{2}]^{1/2},
\label{OmegaI_Gmumu_Omegamunu_bound_general}
\end{eqnarray}
owing to Eq.~(\ref{bounds_gmumu_sum_prod}), where $A_{\rm cell}$ and $V_{\rm cell}$ are the area and volume of the unit cell, respectively. Thus provided the Berry curvature is nonzero, the integration of Eq.~(\ref{bounds_gmumu_sum_prod}) over the momentum space sets a limit on the gauge-invariant part of the spread $\Omega_{I}$.

Owing to the connection between the spread $\Omega_{I}$ and optical effects, Eq.~(\ref{bounds_gmumu_sum_prod}) also constrains optical sum rules of 2D and 3D semiconductors at temperature sufficiently lower than the band gap. For 2D semiconductors, denoting $\overline{\cal O}(\omega)$ as the absorbance under unpolarized light with frequency $\omega$, a relation between $\Omega_{I}$ and the absorbance\cite{CardenasCastillo24_spread_Wannier}
\begin{eqnarray}
&&\Omega_{I}^{2D}=\int_{0}^{\infty}d\omega\,\frac{A_{\rm cell}}{2\pi\omega}\left[\frac{\overline{{\cal O}}(\omega)}{\pi\alpha}\right]\geq A_{\rm cell}\int\frac{d^{2}{\bf k}}{(2\pi)^{2}}|\Omega_{xy}|,
\nonumber \\
\label{OmegaI_opacity_bound}
\end{eqnarray}
where $\alpha\approx 1/137$ is the fine-structure constant, indicates that the absorbance divided by frequency and then integrated over frequency is bounded by the momentum integration of $|\Omega_{xy}|$ according to Eq.~(\ref{OmegaI_Gmumu_Omegamunu_bound_general}). Thus the absorbance measured in any 2D semiconductors that have nonzero Berry curvature, such as transition metal dichalcogenides, is a consequence of this constraint. For 3D semiconductors, the frequency integration of the imaginary part of dielectric function $\varepsilon_{2}(\omega)$ is related to the spread by\cite{CardenasCastillo24_spread_Wannier}
\begin{eqnarray}
&&\Omega_{I}^{3D}=\frac{3V_{\rm cell}\varepsilon_{0}}{\pi e^{2}}\int_{0}^{\infty}d(\hbar\omega)\,\varepsilon_{2}(\omega)
\nonumber \\
&&\;\;\;\;\;\;\;\;\geq\frac{V_{\rm cell}}{\hbar}\int\frac{d^{3}{\bf k}}{(2\pi)^{3}}[\Omega_{xy}^{2}+\Omega_{yz}^{2}+\Omega_{zx}^{2}]^{1/2},
\label{OmegaI_epsilon2_bound}
\end{eqnarray}
and hence the frequency integration of $\varepsilon_{2}(\omega)$ in 3D is also bounded by the momentum integration of $[\Omega_{xy}^{2}+\Omega_{yz}^{2}+\Omega_{zx}^{2}]^{1/2}$. We will use theoretical models of concrete materials to demonstrate these bounds on $\Omega_{I}$ and optical sum rules in the following sections.

To demonstrate the uncertainty relations in solids, we consider finite temperature and treat each momentum ${\bf k}$ as an independent ensemble described by density matrix $\rho({\bf k})=e^{-\beta H({\bf k})}/{\rm Tr}\,e^{-\beta H({\bf k})}$, where $H({\bf k})=\sum_{\ell}\varepsilon_{\ell}c_{\ell}^{\dag}c_{\ell}$ is the second-quantized Hamiltonian, with $\varepsilon_{\ell}$ and $c_{\ell}$ the eigenenergy and electron annihilation operator of band $\ell$ at momentum ${\bf k}$, respectively. The operator ${\cal A}_{\mu}$ under question is expressed in the second-quantized form as
\begin{eqnarray}
{\cal A}_{\mu}=\sum_{\ell\ell '}\langle\ell|{\hat{\cal A}}_{\mu}|\ell '\rangle c_{\ell}^{\dag}c_{\ell '},
\end{eqnarray}
where ${\hat{\cal A}}_{\mu}$ is the single-particle operator acting on the eigenstates. The ensemble averages are then given by
\begin{eqnarray}
&&\langle{\cal A}_{\mu}\rangle=\sum_{\ell}\langle\ell|{\hat{\cal A}}_{\mu}|\ell\rangle f_{\ell},
\nonumber \\
&&\langle{\cal A}_{\mu}{\cal A}_{\nu}\rangle=\sum_{\ell\ell '}\langle\ell|{\hat{\cal A}}_{\mu}|\ell\rangle\langle\ell '|{\hat{\cal A}}_{\nu}|\ell '\rangle f_{\ell}f_{\ell '}
\nonumber \\
&&\qquad\;\;\;\;\;\;+\sum_{\ell\ell '}\langle\ell|{\hat{\cal A}}_{\mu}|\ell '\rangle\langle\ell '|{\hat{\cal A}}_{\nu}|\ell\rangle f_{\ell}(1-f_{\ell '}),
\label{AmuAnu_covariance}
\end{eqnarray}
where $f_{\ell}$ is the Fermi distribution function of band $\ell$ at momentum ${\bf k}$, which can be used to calculate the covariances and commutators that enter the uncertainty relations in Sec.~\ref{sec:uncertainty_mixed_states} and \ref{sec:Efimov_Dodonov_argument}.


\subsection{Optical sum rules for velocity variance}

By choosing the single-particle operator under question to be the group velocity $\hat{\cal A}_{\mu}=v_{\mu}=\partial_{\mu}H$, where $\partial_{\mu}=\partial/\partial k_{\mu}$ is the derivative over momentum, and limiting our discussion to zero temperature such that the Fermi functions satisfy $f_{n}=1$ and $f_{m}=0$, Eq.~(\ref{AmuAnu_covariance}) gives the velocity variance at momentum ${\bf k}$
\begin{eqnarray}
\langle\Delta v_{\mu}^{2}\rangle=\sum_{nm}\langle n|v_{\mu}|m\rangle\langle m|v_{\mu}|n\rangle.
\end{eqnarray}
Combining this result with the optical conductivity at polarization $\mu$ and frequency $\omega$ at momentum ${\bf k}$ 
\begin{eqnarray}
\sigma_{\mu\mu}({\bf k},\omega)=\frac{\pi e^{2}}{a^{D}\hbar\omega}\sum_{nm}\langle n|v_{\mu}|m\rangle\langle m|v_{\mu}|n\rangle\delta(\omega+\frac{\varepsilon_{n}}{\hbar}-\frac{\varepsilon_{m}}{\hbar}),
\nonumber \\
\label{sigmakw_vmuvmu_delta}
\end{eqnarray}
where $a^{D}$ is the volume of the unit cell, one sees that the velocity variance at ${\bf k}$ is given by
\begin{eqnarray}
\langle\Delta v_{\mu}^{2}\rangle=\frac{\hbar a^{D}}{\pi e^{2}}\int d\omega\cdot\omega\cdot\sigma_{\mu\mu}({\bf k},\omega).
\end{eqnarray}
The real part of optical conductivity $\sigma_{1}(\omega)$ measured in real space is that in Eq.~(\ref{sigmakw_vmuvmu_delta}) integrated over ${\bf k}$ and averaged over $D$ directions, so we define a momentum- and direction-averaged velocity variance accordingly
\begin{eqnarray}
&&\frac{1}{D}\sum_{\mu=1}^{D}\overline{\langle\Delta v_{\mu}^{2}\rangle}
\equiv\frac{1}{D}\sum_{\mu=1}^{D}\int\frac{d^{D}{\bf k}}{(2\pi\hbar/a)^{D}}\langle\Delta v_{\mu}^{2}\rangle
\nonumber \\
&&=\frac{\hbar a^{D}}{\pi e^{2}}\int d\omega\frac{\omega}{D}\sum_{\mu=1}^{D}\int\frac{d^{D}{\bf k}}{(2\pi\hbar/a)^{D}}\sigma_{\mu\mu}({\bf k},\omega)
\nonumber \\
&&=\frac{\hbar a^{D}}{\pi e^{2}}\int d\omega\cdot\omega\cdot\sigma_{1}(\omega),
\end{eqnarray}
which indicates an optical sum rule for the velocity variance.

For 3D semiconductors, the relation between the imaginary part of dielectric function and optical conductivity $\varepsilon_{2}(\omega)=\sigma_{1}(\omega)/\varepsilon_{0}\omega$ renders the optical sum rule for the velocity variance
\begin{eqnarray}
&&\frac{1}{3}\sum_{\mu=1}^{3}\overline{\langle\Delta v_{\mu}^{2}\rangle}
=\frac{\hbar V_{\rm cell}\varepsilon_{0}}{\pi e^{2}}\int d\omega\cdot\omega^{2}\cdot\varepsilon_{2}(\omega)
\nonumber \\
&&\geq\frac{1}{3}\int\frac{d^{3}{\bf k}}{(2\pi\hbar/a)^{3}}
\nonumber \\
&&\times\left[\left|\langle[v_{x},v_{y}]\rangle\right|^{2}+\left|\langle[v_{y},v_{z}]\rangle\right|^{2}
+\left|\langle[v_{z},v_{x}]\rangle\right|^{2}\right]^{1/2},\;\;\;
\label{velocity_variance_3D_bound}
\end{eqnarray}
where $V_{\rm cell}$ is the volume of the unit cell. This result indicates that the imaginary part of dielectric function multiplied by frequency square and then integrated over frequency gives the velocity variance of electrons in 3D semiconductors, providing a concrete experimental protocol to measure the variance. In addition, in Eq.~(\ref{velocity_variance_3D_bound}) we have invoked Eq.~(\ref{sum_bound_3D_commutator}) to elaborate that the variance is bounded by the commutators of the velocity operator integrated over momentum, rendering a bound on the optical sum rule.

For 2D semiconductors, using the relation $\sigma_{1}(\omega)=e^{2}\overline{\cal O}(\omega)/4\hbar\pi\alpha$ between the optical conductivity and absorbance $\overline{\cal O}(\omega)$\cite{CardenasCastillo24_spread_Wannier}, where $\alpha\approx 1/137$ is the fine-structure constant, we arrive at the optical sum rule for the velocity variance 
\begin{eqnarray}
&&\frac{1}{2}\sum_{\mu=1}^{2}\overline{\langle\Delta v_{\mu}^{2}\rangle}
=\frac{A_{\rm cell}}{4\pi}\int d\omega\cdot\omega\left[\frac{\overline{\cal O}(\omega)}{\pi\alpha}\right]
\nonumber \\
&&\geq\frac{1}{2}\int\frac{d^{2}{\bf k}}{(2\pi\hbar/a)^{2}}|\langle[v_{x},v_{y}]\rangle|,
\label{2D_velocity_variance_bound_vxvy}
\end{eqnarray}
where $A_{\rm cell}$ is the area of the unit cell. As a result, the absorbance multiplied by frequency and then integrated over frequency gives the velocity variance in 2D semiconductors, bringing its measurement to reality. Moreover, this optical sum rule is bounded by the momentum integration of the commutator of the velocity operators according to the uncertainty relation in Eq.~(\ref{sum_bound_2D_commutator}).

The significance of these results is that, contrary to the common wisdom that insulators and semiconductors lack a characteristic velocity scale like the Fermi velocity in metals, the (square root of) velocity variance actually serves as a characteristic velocity scale that determines the optical properties of semiconductors. Moreover, $\omega^{2}\varepsilon_{2}(\omega)$ in 3D and $\omega\,\overline{\cal O}(\omega)$ in 2D serve as spectral functions for the velocity variance. In the following sections, we will use lattice models of 2D and 3D semiconductors to estimate the velocity variance for concrete materials.

\begin{figure}[ht]
\begin{center}
\includegraphics[clip=true,width=0.99\columnwidth]{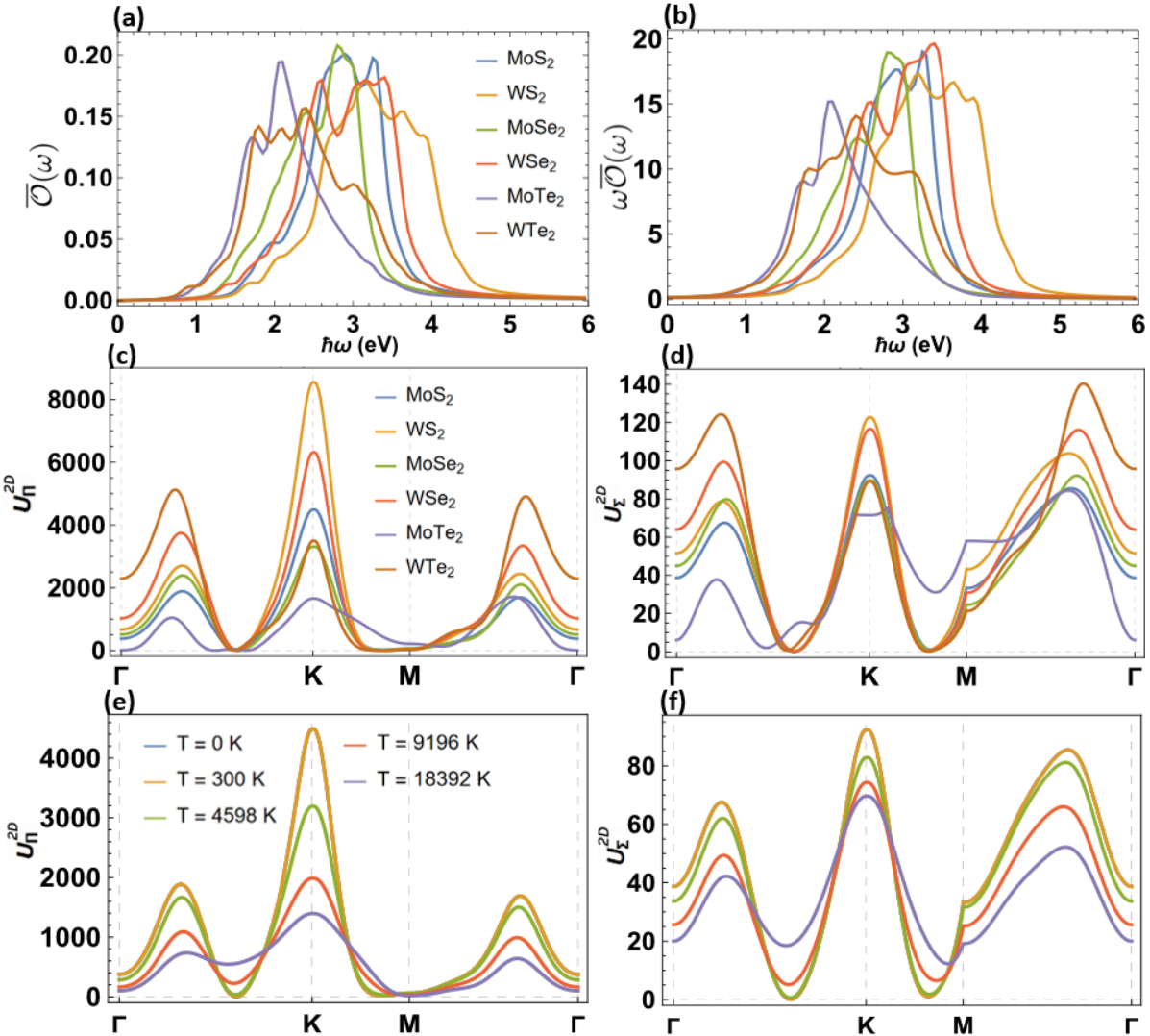}
\caption{(a) Absorbance $\overline{\cal O}(\omega)$ and (b) frequency times absorbance $\omega\overline{\cal O}(\omega)$ at zero temperature for the six TMD materials under investigation. (c) The product bound $U_{\prod}^{2D}$ and (d) summation bound $U_{\Sigma}^{2D}$ at zero temperature plotted along the high-symmetry line $\Gamma-K-M-\Gamma$ for the six materials. (e) $U_{\prod}^{2D}$ and (f) $U_{\Sigma}^{2D}$ at finite temperature for MoS$_{2}$. } 
\label{fig:TMD_figure}
\end{center}
\end{figure}

\begin{table}[htbp]
    \centering
    \caption{
    Unit-cell area $A_{\mathrm{cell}}$, spread of Wannier function and its lower bound
    $\Omega_I\geq t_{1}\equiv A_{\mathrm{cell}}\overline{|\Omega_{xy}|}\,(\mathrm{\AA}^2)$, and the two sides of the
    Brillouin-zone-integrated velocity-fluctuation inequality $t_{2}\equiv \frac{1}{2}\sum_{\mu}\overline{\langle\Delta v_{\mu}^2\rangle}\,
          (\mathrm{eV\AA}/\hbar)^2 \geq t_{3}\equiv \frac{1}{2}\overline{|\langle[v_x,v_y]\rangle|}\,
          (\mathrm{eV\AA}/\hbar)^2$ in Eq.~(\ref{2D_velocity_variance_bound_vxvy}) for the six monolayer TMDs at $T=0$ K.
    }
    \label{tab:velocity_fluctuation}

    \begin{tabular}{cccccc}
        \toprule
        \textbf{Mat.}
        & $A_{\mathrm{cell}}\,(\mathrm{\AA}^2)$
        & $\Omega_I\,(\mathrm{\AA}^2)$
        & $t_{1}$
        & $t_{2}$
        & $t_{3}$
        \\
        \midrule

        MoS$_2$  & 8.813  & 5.651 & 3.792 & 21.886 & 14.104 \\
        WS$_2$   & 8.818  & 5.738 & 3.842 & 28.695 & 19.116 \\
        MoSe$_2$ & 9.580  & 6.522 & 4.345 & 20.206 & 12.354 \\
        WSe$_2$  & 9.574  & 6.786 & 4.490 & 26.697 & 16.836 \\
        MoTe$_2$ & 10.957 & 8.137 & 4.267 & 17.415 & 9.058  \\
        WTe$_2$  & 10.976 & 8.773 & 5.800 & 22.630 & 12.513 \\

        \bottomrule
    \end{tabular}
\end{table}

\subsection{Application: 2D transition metal dichalcogenides \label{sec:TMD}}

In this section, we use 2D transition metal dichalcogenides $1H-MX_{2}$ to demonstrate the optical sum rules for the velocity variance and spread of Wannier function, and their bounds caused by Berry curvature. The six-band effective model for three transition metal (Mo,W) $d$ orbitals and their two spin projections has the basis
\begin{eqnarray}
|\psi\rangle=(d_{z^{2}}\uparrow,d_{xy}\uparrow,d_{x^{2}-y^{2}}\uparrow,
d_{z^{2}}\downarrow,|d_{xy}\downarrow,d_{x^{2}-y^{2}}\downarrow).
\nonumber \\
\end{eqnarray}
The orbital hoppings for each spin sector are described by the $3\times 3$ Hamiltonian
\begin{eqnarray}
H_{0}({\bf k})=\left(\begin{array}{ccc}
V_{0} & V_{1} & V_{2} \\
V_{1}^{\ast} & V_{11} & V_{12} \\
V_{2}^{\ast} & V_{12}^{\ast} & V_{22}
\end{array}\right)
\end{eqnarray}
whose ${\bf k}$-dependence for the harmonics $V_{0}\sim V_{22}$ and parameters for each TMD compound are given in Ref.~\onlinecite{Liu13}. The full $6\times 6$ Hamiltonian that includes the spin-orbit coupling (SOC) $\lambda$ takes the form
\begin{eqnarray}
&&H({\bf k})=\left(\begin{array}{cc}
H_{0}({\bf k})+\frac{\lambda}{2}L_{z} & 0 \\
0 & H_{0}({\bf k})-\frac{\lambda}{2}L_{z}
\end{array}\right),
\nonumber \\
&&L_{z}=\left(\begin{array}{ccc}
0 & 0 & 0 \\
0 & 0 & 2i \\
0 & -2i & 0
\end{array}\right).
\end{eqnarray} 
The quantum metric and Berry curvature of this model has been addressed previously\cite{Liu13,CardenasCastillo24_spread_Wannier}, and the gauge-invariant part of the spread of Wannier function $\Omega_{I}$ has been extracted, as listed in Table \ref{tab:velocity_fluctuation} (a factor of 3 is corrected from Ref.~\onlinecite{CardenasCastillo24_spread_Wannier}). Numerical calculation for the bound on $\Omega_{I}$ caused by momentum integration of the modulus of Berry curvature $A_{\mathrm{cell}}\overline{|\Omega_{xy}|}$ in Eq.~(\ref{OmegaI_opacity_bound}) is given in Table \ref{tab:velocity_fluctuation}. Remarkably, the ratio between the spread and its bound remains roughly the same $\Omega_{I}/A_{\mathrm{cell}}\overline{|\Omega_{xy}|}\approx 1.5$ for five out of the six materials examined (with the exception of MoTe$_{2}$), indicating a very strong constraint that the Berry curvature imposes on the spread. For the velocity variance $\frac{1}{2}\sum_{\mu}\overline{\langle\Delta v_{\mu}^2\rangle}$ and its bound $\frac{1}{2}\overline{|\langle[v_x,v_y]\rangle|}$, described by Eq.~(\ref{2D_velocity_variance_bound_vxvy}), the values in Table \ref{tab:velocity_fluctuation} give a ratio of about 1.5$\sim$1.9, which also suggests a very strong constraint that the commutator of velocities $[v_{x},v_{y}]$ imposes on the velocity variance. In Fig.~\ref{fig:TMD_figure} (a), we present the absorbance $\overline{\cal O}(\omega)$ of the six TMD materials calculated by this tight-binding model\cite{CardenasCastillo24_spread_Wannier}. Figure \ref{fig:TMD_figure} (b) further shows the profile of frequency times absorbance $\omega\cdot\overline{\cal O}(\omega)$ for the six materials, whose frequency integration (area underneath the curves) gives the velocity variance according to Eq.~(\ref{2D_velocity_variance_bound_vxvy}), demonstrating the simple experimental protocol to extract the velocity variance. 


We proceed to use velocity operator $\hat{\cal A}_{\mu}=v_{\mu}$ to demonstrate that the uncertainty relations are satisfied at any momentum. In Fig.~\ref{fig:TMD_figure} (c), we present $U_{\Pi}^{2D}$ as the left hand side minus the right hand side of the Robertson-Schr\"{o}dinger uncertainty relation in Eq.~(\ref{Robertson_Schrodinger_Heisenberg_mixed_states}) plotted along the high-symmetry line $\Gamma-K-M-\Gamma$, and similarly in Fig.~\ref{fig:TMD_figure} (d) $U_{\Sigma}^{2D}$ as the left hand side minus the right hand side of the uncertainty relation in Eq.~(\ref{sum_bound_2D_commutator}). For all six materials at zero temperature, $U_{\Pi}^{2D}\geq 0$ and $U_{\Sigma}^{2D}\geq 0$ is satisfied at any momentum, indicating the validity of the uncertainty relations. In Fig.~(e) and (f), we choose MoS$_{2}$ to examine $U_{\Pi}^{2D}$ and $U_{\Sigma}^{2D}$ at finite temperature, where one sees that even as temperature gradually increases, $U_{\Pi}^{2D}\geq 0$ and $U_{\Sigma}^{2D}\geq 0$ are still satisfied, confirming the validity of Eqs.~(\ref{Robertson_Schrodinger_Heisenberg_mixed_states}) and (\ref{sum_bound_2D_commutator}) even at finite temperature (we have examined up to an unreasonably high temperature $T=18392$K comparable to the band gap of MoS$_{2}$ for the sake of demonstration, even though in reality the material would have melted at this temperature). 

\subsection{Application: 3D topological semimetals}

3D topological semimetals serve as examples to analytically demonstrate the 3D bounds in Eq.~(\ref{bounds_gmumu_sum_prod}). The low energy sector of type-I Weyl semimetals, like TaAs and TaP\cite{Weng15,Huang15_2}, contains pairs of Weyl points\cite{Yan17,Armitage18,Lv21}. The Hamiltonian per spin species near each valley is well described by the $2\times 2$ Dirac model 
\begin{eqnarray}
H({\bf k})=\pm\left(vk_{x}\sigma_{x}+vk_{y}\sigma_{y}+vk_{z}\sigma_{z}\right)={\bf d}\cdot{\boldsymbol\sigma},
\label{Weyl_Hamiltonian}
\end{eqnarray}
where the $\pm$ sign characterizes the chirality of each Weyl point. In the Cartesian coordinates, the quantum metric and Berry curvature obtained from Eq.~(\ref{gmunu_Omegamunu_nm_expression}) are
\begin{eqnarray}
&&g_{\mu\mu}=\frac{1}{4k^{4}}\left(k^{2}-k_{\mu}^{2}\right),\;\;\;
g_{\mu\neq\nu}=-\frac{k_{\mu}k_{\nu}}{4k^{4}},
\nonumber \\
&&\Omega_{xy}=\pm\frac{k_{z}}{2k^{3}},
\label{Weyl_semimetal_gmunu_Omegamunu}
\end{eqnarray}
and other components of Berry curvature can be obtained from the cyclic replacement of the three indices. At zero temperature, the bound on the summation of quantum metric in Eq.~(\ref{bounds_gmumu_sum_prod}) is saturated by 
\begin{eqnarray}
g_{xx}+g_{yy}+g_{zz}=\left[\Omega_{xy}^{2}+\Omega_{yz}^{2}+\Omega_{zx}^{2}\right]^{1/2}=\frac{1}{2k^{2}},
\end{eqnarray}
and the third inequality in Eq.~(\ref{bounds_gmumu_sum_prod}) is satisfied by
\begin{eqnarray}
&&g_{xx}g_{yy}g_{zz}-\frac{1}{9}\left(g_{xx}\Omega_{yz}^{2}+g_{yy}\Omega_{zx}^{2}+g_{zz}\Omega_{xy}^{2}\right)
\nonumber \\
&&=\frac{1}{576 k^{12}}\left\{(k_{x}^{2}k_{y}^{2}+k_{y}^{2}k_{z}^{2}+k_{z}^{2}k_{x}^{2})k^{2}-9k_{x}^{2}k_{y}^{2}k_{z}^{2}\right\}\geq 0,
\nonumber \\
\end{eqnarray}
due to the AM-GM inequality $k_{x}^{4}k_{y}^{2}+k_{y}^{2}k_{z}^{4}\geq 2\sqrt{k_{x}^{4}k_{y}^{4}k_{z}^{4}}$. 


Now we choose the positive chirality and consider the velocity operator $\hat{\cal A}_{\mu}=\partial H/\partial k_{\mu}=v\sigma_{\mu}$, and define
\begin{eqnarray}
&&\varepsilon_{\gamma}=\gamma vk,\qquad f_{\gamma}=\frac{1}{e^{\beta\varepsilon_{\gamma}}+1},\qquad n_{\mu}=\frac{k_{\mu}}{k},
\nonumber \\
&&a=1-2f_{+}f_{-},\qquad b=(f_{+}-f_{-})^{2},
\end{eqnarray}
where $\gamma=\pm$. The covariance and commutator in Eq.~(\ref{Kmunu_Dmunu_Bmunu_A}) for the velocity operators are found to be 
\[
\begin{aligned}
C_{\mu\nu} &= v^2\left(a\delta_{\mu\nu}-bn_\mu n_\nu\right),\\
B_{\mu\nu} &= -2\gamma v^2(f_+-f_-)
\epsilon_{\mu\nu\lambda}n_\lambda,
\end{aligned}
\]
from which we can analytically examine the uncertainty relations in Sec.~\ref{sec:uncertainty_mixed_states} and \ref{sec:Efimov_Dodonov_argument}. The single-observable uncertainty relations in Eq.~(\ref{QCRB_pure_state_3C_uncertainty_interpretation}) are satisfied by 
\begin{eqnarray}
\langle\Delta v_{\mu}^{2}\rangle= v^2\left(a-bn_\mu^2\right)
\geq v^2\frac{b}{a}\left(1-n_\mu^2\right),
\end{eqnarray}
since $a^{2}-b=(1-a)^{2}\geq 0$, and the bounds are saturated at zero temperature when $a=b=1$. The summation bound in Eq.~(\ref{sum_bound_3D_commutator}) takes the form
\begin{eqnarray}
2v^{2}(1-f_{+}f_{-})\geq 2v^{2}(f_{-}-f_{+}),
\end{eqnarray}
which is always satisfied, and it saturates at zero temperature. The product bound in Eq.~(\ref{product_bound_3D_commutator}) requires
\begin{eqnarray}
\prod_{\mu}\left(a-bn_\mu^2\right)\geq \frac{4b}{9}\left(a-b\sum_\mu n_\mu^4\right),
\label{product_bound_3D_semimetal_formula}
\end{eqnarray}
which is also satisfied.


\begin{table}[t]
\centering
\caption{The gauge-invariant part of the spread of Wannier function $\Omega_{I}$, the bound on the spread $V_{\Omega_{I}}$, the velocity variance $\overline{\Delta v^2}$ in units of $10^{12}\,\mathrm{m^2/s^2}$, and the bound on the velocity variance $V_{\overline{\Delta v^2}}$ in the same unit for ten semiconductors obtained from the $sp^3s^*$ model.}
\label{tab:bounds_semiconductors}

\small
\setlength{\tabcolsep}{6pt}
\renewcommand{\arraystretch}{1.15}

\begin{tabular}{lcccc}
\hline
Mat. 
& $\Omega_I\;(\mathrm{\AA}^2)$ 
& $V_{\Omega_I}\;(\mathrm{\AA}^2)$
& $\overline{\Delta v^2}$
& $V_{\overline{\Delta v^2}}$
\\
\hline
C    & 5.5352  & 0 & 23.9992 & 0 \\
Si   & 13.3088 & 0 & 9.0492  & 0 \\
Ge   & 16.1844 & 0 & 8.7846  & 0 \\
SiC  & 7.9746  & 0.4714      & 12.3333 & 0.2100 \\
GaP  & 11.8672 & 0.5903      & 8.5913  & 0.1356 \\
GaAs & 13.6271 & 0.7526      & 8.4343  & 0.2000 \\
GaSb & 17.2009 & 0.9899      & 7.7078  & 0.2402 \\
InP  & 13.1939 & 0.8326      & 6.1185  & 0.1608 \\
InAs & 15.8405 & 0.9877      & 6.6188  & 0.1593 \\
InSb & 19.4416 & 1.2381      & 6.3807  & 0.1827 \\
\hline
\end{tabular}
\end{table}

\subsection{Application: 3D common semiconductors}

In this section, we use the $sp^{3}s^{\ast}$ tight-binding model to corroborate the geometric bounds and uncertainty relations for common group IV and group III-V semiconductors. This tight-binding model is known to give a reasonable description for the low-energy band structure relevant to optical properties of these materials, such as the dielectric function $\varepsilon_{2}(\omega)$. Details of the 
model and the calculation of optical properties can be found in Ref.~\onlinecite{Porlles26_semiconductor_metric}, so we omit them for simplicity. The spread of Wannier function $\Omega_{I}$ (a factor of 4 is corrected from Ref.~\onlinecite{Porlles26_semiconductor_metric}), the bound on the spread $V_{\Omega_{I}}$ described by the right hand side of Eq.~(\ref{OmegaI_epsilon2_bound}), the velocity variance $\overline{\Delta v^2}$, and the bound on velocity variance $V_{\overline{\Delta v^2}}$ described by the right hand side of Eq.~(\ref{velocity_variance_3D_bound}) calculated for 10 common semiconductors are listed in Table \ref{tab:bounds_semiconductors}. The absolute scale of the velocity variance $\overline{\Delta v^2}$ is found to be roughly $\sim 10^{12}$m$^{2}/$s$^{2}$ for these materials, meaning that the standard deviation of velocity $(\overline{\Delta v^2})^{1/2}$ is roughly $\sim 10^{6}$m/s, comparable to the typical value of Fermi velocity in metals. On the other hand, the bound on the spread of Wannier function $\Omega_{I}$ is found to be an order of magnitude smaller than the spread itself $V_{\Omega_I}\ll\Omega_{I}$, suggesting a very small Berry curvature in group III-V semiconductors (see the discussion below). Similarly, the bound on the velocity variance is also found to be more than an order of magnitude smaller $V_{\overline{\Delta v^2}}\ll \overline{\Delta v^2}$, suggesting a small commutator $\langle[v_{\mu},v_{\nu}]\rangle$.

\begin{figure}[ht]
\begin{center}
\includegraphics[clip=true,width=0.99\columnwidth]{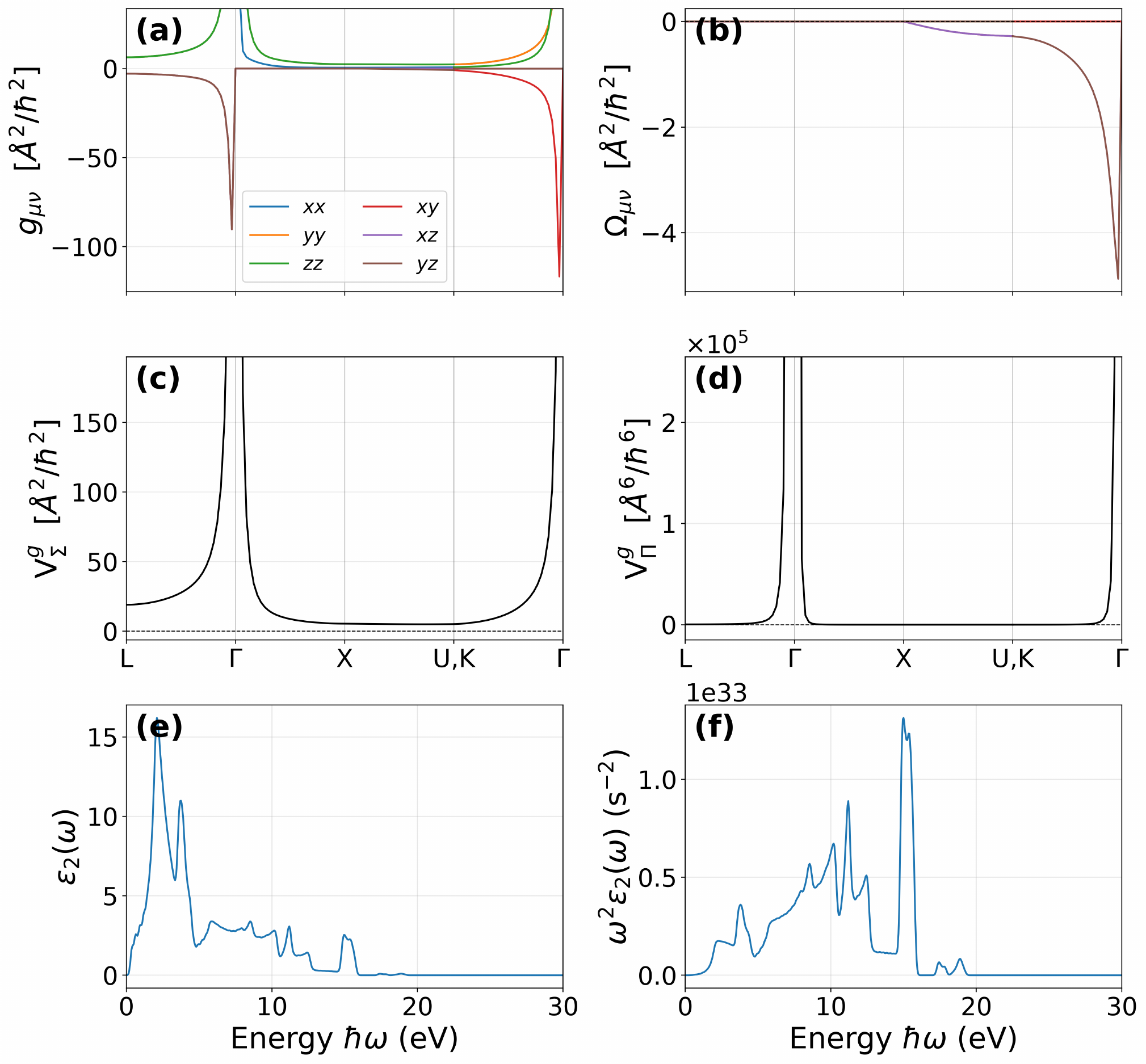}
\caption{Numerical results for InSb: (a) Quantum metric and (b) Berry curvature along the high-symmetry line. (c) and (d) show the left hand sides minus the right hand sides of the first and the third inequalities in Eq.~(\ref{bounds_gmumu_sum_prod}). (e) The dielectric function $\varepsilon_{2}(\omega)$ that integrates to the spread $\Omega_{I}$, and (f) frequency square times the dielectric function $\omega^{2}\cdot\varepsilon_{2}(\omega)$ that integrates to the velocity variance $\overline{\Delta v^2}$. } 
\label{fig:Geometric_bounds_InSb}
\end{center}
\end{figure}

To elaborate the geometric bounds at each momentum ${\bf k}$, note that group IV semiconductors (C, Si, Ge) have the diamond structure that preserves inversion symmetry, rendering a vanishing Berry curvature $\Omega_{\mu\nu}=0$ everywhere in the Brillouin zone, so the geometric bounds in Eq.~(\ref{bounds_gmumu_sum_prod}) are trivially satisfied. In contrast, group III-V semiconductors have the zinc-blende structure that breaks inversion symmetry, so the Berry curvature is nonzero. As an example, the results for InSb are shown in Fig.~\ref{fig:Geometric_bounds_InSb} (a) and (b) along the high-symmetry line $L-\Gamma-X-(U,K)-\Gamma$, which manifest a very singular profile of quantum metric and Berry curvature owing to the small band gap at the $\Gamma$-point. The quantum metric is found to be two orders of magnitude larger than Berry curvature, so the summation bound $V_{\Sigma}^{g}$ and the product bound $V_{\Pi}^{g}$ shown in Fig.~\ref{fig:Geometric_bounds_InSb} (c) and (d), defined as the left hand sides minus the right hand sides of the first and the third inequalities in Eq.~(\ref{bounds_gmumu_sum_prod}), are only weakly constrained by the Berry curvature. Finally, to elaborate the proposed experimental protocol, the dielectric function $\varepsilon_{2}(\omega)$ of InSb calculated from the $sp^{3}s^{\ast}$ model is presented in Fig.~\ref{fig:Geometric_bounds_InSb} (e), whose frequency integration gives the spread of Wannier function $\Omega_{I}$ according to Eq.~(\ref{OmegaI_epsilon2_bound}). Similarly, the frequency square times the dielectric function $\omega^{2}\cdot\varepsilon_{2}(\omega)$ of InSb is shown in Fig.~\ref{fig:Geometric_bounds_InSb} (f), whose frequency integration extracts the velocity variance according to Eq.~(\ref{velocity_variance_3D_bound}).


\begin{figure}[ht]
\begin{center}
\includegraphics[clip=true,width=0.99\columnwidth]{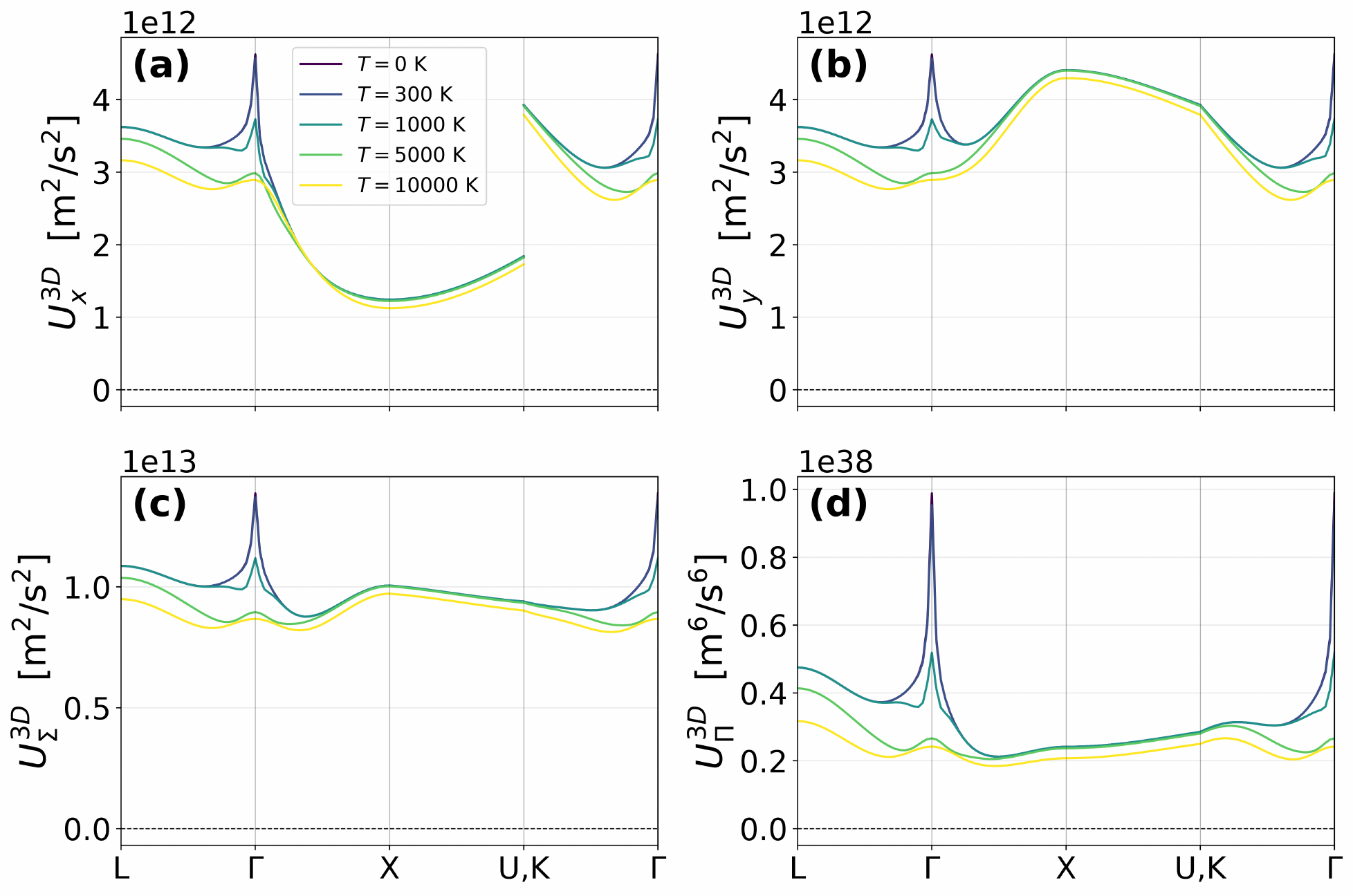}
\caption{Uncertainty relations for the velocity operator ${\cal A}_{\mu}=v_{\mu}$ in InSb plotted along a high-symmetry line at different temperatures: Single-observable uncertainty relations (a) $U_{x}^{3D}$ and (b) $U_{y}^{3D}$ in Eq.~(\ref{QCRB_pure_state_3C_uncertainty_interpretation}), (c) the summation bound $U_{\Sigma}^{3D}$ in Eq.~(\ref{sum_bound_3D_commutator}) and (d) and product bound $U_{\Pi}^{3D}$ in Eq.~(\ref{product_bound_3D_commutator}). } 
\label{fig:Uncertainty_relations_InSb}
\end{center}
\end{figure}

Similar to Sec.~\ref{sec:TMD}, we use the velocity operator $\hat{\cal A}_{\mu}=v_{\mu}$ to demonstrate the uncertainty relations in 3D. In Fig.~\ref{fig:Uncertainty_relations_InSb} (a) and (b), we present the single-observable uncertainty relations $U_{x}^{3D}$ and $U_{y}^{3D}$ defined as the left hand sides minus the right hand sides of the first two inequalities of Eq.~(\ref{QCRB_pure_state_3C_uncertainty_interpretation}), whose positiveness at any  momentum and any temperature indicates that they are always satisfied. In Fig.~\ref{fig:Uncertainty_relations_InSb} (c) and (d), the summation bound $U_{\Sigma}^{3D}$ for Eq.~(\ref{sum_bound_3D_commutator}) and product bound $U_{\Pi}^{3D}$ for Eq.~(\ref{product_bound_3D_commutator}), defined as the left hand side minus the right hand side of these inequalities, are presented, whose positiveness at any momentum indicates the validity of these uncertainty relations at any momentum and temperature.

\section{Conclusions}

In summary, several uncertainty relations are derived for the diagonal product, trace, and each diagonal element of the covariance matrix of multiple observables in mixed states. These uncertainty relations imply geometric bounds on QFIM caused by MUC in mixed states, which in pure states become geometric bounds on quantum metric caused by Berry curvature. As a result, the spread of valence-band Wannier function is shown to be bounded by Berry curvature, which also bounds the frequency integrations of dielectric function in 3D and absorbance divided by frequency in 2D. Our formalism further reveals the velocity variance as a characteristic velocity scale that determines the optical sum rules of frequency square times dielectric function in 3D 
and frequency times absorbance in 2D, thereby challenges the conventional view that semiconductors and insulators lack a characteristic velocity scale. Moreover, the velocity variance and optical sum rules are found to be bounded by the commutators of 
velocity operators. As demonstrated by prototype semiconductors and topological materials, we anticipate that the uncertainty relations, geometric bounds, velocity variance, and optical sum rules revealed by the present work have a wide application in solids and other mixed states, a prospect we leave for future investigations.

\begin{acknowledgments}

We acknowledge the support of the INCT project Advanced Quantum Materials, 
involving the Brazilian agencies CNPq (Proc. 408766/2024-7), FAPESP, and CAPES. W.C. acknowledges the financial support of the productivity in research fellowship from CNPq. 

\end{acknowledgments}

\appendix

\section{Index-explicit inequalities in 3D \label{apx:index_explicity_inequalities_3D}}

In this section, we list the index-explicit formulas for the uncertainty relation in Eq.~(\ref{selfbound_uncertainty_interpretation_generator}) for 3D cases $\mu=\left\{x,y,z\right\}$, which are completely analogous to the pure state results\cite{Chen26_QCRB}
\begin{widetext}
\begin{eqnarray}
&&\langle\Delta{\cal A}_{x}^{2}\rangle\geq\frac{1}{4\det C_{\mu\nu}}\left(-\langle[{\cal A}_{x},{\cal A}_{y}]\rangle^{2}\langle\Delta{\cal A}_{z}^{2}\rangle\langle\Delta{\cal A}_{x}^{2}\rangle
+\frac{1}{4}\langle[{\cal A}_{x},{\cal A}_{y}]\rangle^{2}\langle\left\{\Delta{\cal A}_{z},\Delta{\cal A}_{x}\right\}\rangle^{2}\right.
\nonumber \\
&&+\frac{1}{2}\langle[{\cal A}_{x},{\cal A}_{y}]\rangle\langle[{\cal A}_{z},{\cal A}_{x}]\rangle
\langle\left\{\Delta{\cal A}_{x},\Delta{\cal A}_{y}\right\}\rangle\langle\left\{\Delta{\cal A}_{z},\Delta{\cal A}_{x}\right\}\rangle
-\langle[{\cal A}_{x},{\cal A}_{y}]\rangle\langle[{\cal A}_{z},{\cal A}_{x}]\rangle\langle\left\{\Delta{\cal A}_{y},\Delta{\cal A}_{z}\right\}\rangle
\langle\Delta{\cal A}_{x}^{2}\rangle
\nonumber \\
&&\left.-\langle[{\cal A}_{z},{\cal A}_{x}]\rangle^{2}\langle\Delta{\cal A}_{x}^{2}\rangle\langle\Delta{\cal A}_{y}^{2}\rangle
+\frac{1}{4}\langle[{\cal A}_{z},{\cal A}_{x}]\rangle^{2}\langle\left\{\Delta{\cal A}_{x},\Delta{\cal A}_{y}\right\}\rangle^{2}\right),
\nonumber \\
&&\langle\Delta{\cal A}_{y}^{2}\rangle\geq\frac{1}{4\det C_{\mu\nu}}\left(-\langle[{\cal A}_{y},{\cal A}_{z}]\rangle^{2}\langle\Delta{\cal A}_{x}^{2}\rangle\langle\Delta{\cal A}_{y}^{2}\rangle
+\frac{1}{4}\langle[{\cal A}_{y},{\cal A}_{z}]\rangle^{2}\langle\left\{\Delta{\cal A}_{x},\Delta{\cal A}_{y}\right\}\rangle^{2}\right.
\nonumber \\
&&+\frac{1}{2}\langle[{\cal A}_{y},{\cal A}_{z}]\rangle\langle[{\cal A}_{x},{\cal A}_{y}]\rangle
\langle\left\{\Delta{\cal A}_{y},\Delta{\cal A}_{z}\right\}\rangle\langle\left\{\Delta{\cal A}_{x},\Delta{\cal A}_{y}\right\}\rangle
-\langle[{\cal A}_{y},{\cal A}_{z}]\rangle\langle[{\cal A}_{x},{\cal A}_{y}]\rangle\langle\left\{\Delta{\cal A}_{z},\Delta{\cal A}_{x}\right\}\rangle
\langle\Delta{\cal A}_{y}^{2}\rangle
\nonumber \\
&&\left.-\langle[{\cal A}_{x},{\cal A}_{y}]\rangle^{2}\langle\Delta{\cal A}_{y}^{2}\rangle\langle\Delta{\cal A}_{z}^{2}\rangle
+\frac{1}{4}\langle[{\cal A}_{x},{\cal A}_{y}]\rangle^{2}\langle\left\{\Delta{\cal A}_{y},\Delta{\cal A}_{z}\right\}\rangle^{2}\right),
\nonumber \\
&&\langle\Delta{\cal A}_{z}^{2}\rangle\geq\frac{1}{4\det C_{\mu\nu}}\left(-\langle[{\cal A}_{z},{\cal A}_{x}]\rangle^{2}\langle\Delta{\cal A}_{y}^{2}\rangle\langle\Delta{\cal A}_{z}^{2}\rangle
+\frac{1}{4}\langle[{\cal A}_{z},{\cal A}_{x}]\rangle^{2}\langle\left\{\Delta{\cal A}_{y},\Delta{\cal A}_{z}\right\}\rangle^{2}\right.
\nonumber \\
&&+\frac{1}{2}\langle[{\cal A}_{z},{\cal A}_{x}]\rangle\langle[{\cal A}_{y},{\cal A}_{z}]\rangle
\langle\left\{\Delta{\cal A}_{z},\Delta{\cal A}_{x}\right\}\rangle\langle\left\{\Delta{\cal A}_{y},\Delta{\cal A}_{z}\right\}\rangle
-\langle[{\cal A}_{z},{\cal A}_{x}]\rangle\langle[{\cal A}_{y},{\cal A}_{z}]\rangle\langle\left\{\Delta{\cal A}_{x},\Delta{\cal A}_{y}\right\}\rangle
\langle\Delta{\cal A}_{z}^{2}\rangle
\nonumber \\
&&\left.-\langle[{\cal A}_{y},{\cal A}_{z}]\rangle^{2}\langle\Delta{\cal A}_{z}^{2}\rangle\langle\Delta{\cal A}_{x}^{2}\rangle
+\frac{1}{4}\langle[{\cal A}_{y},{\cal A}_{z}]\rangle^{2}\langle\left\{\Delta{\cal A}_{z},\Delta{\cal A}_{x}\right\}\rangle^{2}\right).
\label{QCRB_pure_state_3C_uncertainty_interpretation}
\end{eqnarray}
In particular, by choosing the operator to be the SLD ${\cal A}_{\mu}=L_{\mu}$, these uncertainty relations become geometric bounds 
\begin{eqnarray}
    F_{xx} &\geq& \frac{4\,{\cal U}_{xy}^{2}(F_{zz}F_{xx}-F_{zx}^{2})+4\,{\cal U}_{zx}^{2}(F_{xx}F_{yy}-F_{xy}^{2}) -8\,{\cal U}_{xy}{\cal U}_{zx}(F_{xy}F_{zx}-F_{xx}F_{yz})}{F_{xx}F_{yy}F_{zz}+2F_{xy}F_{yz}F_{zx} -F_{xx}F_{yz}^{2}-F_{yy}F_{zx}^{2}-F_{zz}F_{xy}^{2}},
    \nonumber \\
    F_{yy} &\geq& \frac{4\,{\cal U}_{yz}^{2}(F_{xx}F_{yy}-F_{xy}^{2})+4\,{\cal U}_{xy}^{2}(F_{yy}F_{zz}-F_{yz}^{2}) -8\,{\cal U}_{yz}{\cal U}_{xy}(F_{yz}F_{xy}-F_{yy}F_{zx})}{F_{xx}F_{yy}F_{zz}+2F_{xy}F_{yz}F_{zx} -F_{xx}F_{yz}^{2}-F_{yy}F_{zx}^{2}-F_{zz}F_{xy}^{2}},
    \nonumber \\
    F_{zz} &\geq& \frac{4\,{\cal U}_{zx}^{2}(F_{yy}F_{zz}-F_{yz}^{2})+4\,{\cal U}_{yz}^{2}(F_{zz}F_{xx}-F_{zx}^{2}) -8\,{\cal U}_{zx}{\cal U}_{yz}(F_{zx}F_{yz}-F_{zz}F_{xy})}{F_{xx}F_{yy}F_{zz}+2F_{xy}F_{yz}F_{zx} -F_{xx}F_{yz}^{2}-F_{yy}F_{zx}^{2}-F_{zz}F_{xy}^{2}},
\label{selfbound_3C_index_explicit}
\end{eqnarray}
\end{widetext}
that constrain the diagonal elements of QFIM by MUC in 3D parameter spaces. 


\bibliography{Literatur}

\begin{thebibliography}{56}%
\makeatletter
\providecommand \@ifxundefined [1]{%
 \@ifx{#1\undefined}
}%
\providecommand \@ifnum [1]{%
 \ifnum #1\expandafter \@firstoftwo
 \else \expandafter \@secondoftwo
 \fi
}%
\providecommand \@ifx [1]{%
 \ifx #1\expandafter \@firstoftwo
 \else \expandafter \@secondoftwo
 \fi
}%
\providecommand \natexlab [1]{#1}%
\providecommand \enquote  [1]{``#1''}%
\providecommand \bibnamefont  [1]{#1}%
\providecommand \bibfnamefont [1]{#1}%
\providecommand \citenamefont [1]{#1}%
\providecommand \href@noop [0]{\@secondoftwo}%
\providecommand \href [0]{\begingroup \@sanitize@url \@href}%
\providecommand \@href[1]{\@@startlink{#1}\@@href}%
\providecommand \@@href[1]{\endgroup#1\@@endlink}%
\providecommand \@sanitize@url [0]{\catcode `\\12\catcode `\$12\catcode
  `\&12\catcode `\#12\catcode `\^12\catcode `\_12\catcode `\%12\relax}%
\providecommand \@@startlink[1]{}%
\providecommand \@@endlink[0]{}%
\providecommand \url  [0]{\begingroup\@sanitize@url \@url }%
\providecommand \@url [1]{\endgroup\@href {#1}{\urlprefix }}%
\providecommand \urlprefix  [0]{URL }%
\providecommand \Eprint [0]{\href }%
\providecommand \doibase [0]{https://doi.org/}%
\providecommand \selectlanguage [0]{\@gobble}%
\providecommand \bibinfo  [0]{\@secondoftwo}%
\providecommand \bibfield  [0]{\@secondoftwo}%
\providecommand \translation [1]{[#1]}%
\providecommand \BibitemOpen [0]{}%
\providecommand \bibitemStop [0]{}%
\providecommand \bibitemNoStop [0]{.\EOS\space}%
\providecommand \EOS [0]{\spacefactor3000\relax}%
\providecommand \BibitemShut  [1]{\csname bibitem#1\endcsname}%
\let\auto@bib@innerbib\@empty
\bibitem [{\citenamefont {Ozawa}\ and\ \citenamefont
  {Goldman}(2018)}]{Ozawa18}%
  \BibitemOpen
  \bibfield  {author} {\bibinfo {author} {\bibfnamefont {T.}~\bibnamefont
  {Ozawa}}\ and\ \bibinfo {author} {\bibfnamefont {N.}~\bibnamefont
  {Goldman}},\ }\bibfield  {title} {\bibinfo {title} {Extracting the quantum
  metric tensor through periodic driving},\ }\href
  {https://doi.org/10.1103/PhysRevB.97.201117} {\bibfield  {journal} {\bibinfo
  {journal} {Phys. Rev. B}\ }\textbf {\bibinfo {volume} {97}},\ \bibinfo
  {pages} {201117} (\bibinfo {year} {2018})}\BibitemShut {NoStop}%
\bibitem [{\citenamefont {Ahn}\ \emph {et~al.}(2022)\citenamefont {Ahn},
  \citenamefont {Guo}, \citenamefont {Nagaosa},\ and\ \citenamefont
  {Vishwanath}}]{Ahn22}%
  \BibitemOpen
  \bibfield  {author} {\bibinfo {author} {\bibfnamefont {J.}~\bibnamefont
  {Ahn}}, \bibinfo {author} {\bibfnamefont {G.-Y.}\ \bibnamefont {Guo}},
  \bibinfo {author} {\bibfnamefont {N.}~\bibnamefont {Nagaosa}},\ and\ \bibinfo
  {author} {\bibfnamefont {A.}~\bibnamefont {Vishwanath}},\ }\bibfield  {title}
  {\bibinfo {title} {Riemannian geometry of resonant optical responses},\
  }\href {https://doi.org/10.1038/s41567-021-01465-z} {\bibfield  {journal}
  {\bibinfo  {journal} {Nat. Phys.}\ }\textbf {\bibinfo {volume} {18}},\
  \bibinfo {pages} {290} (\bibinfo {year} {2022})}\BibitemShut {NoStop}%
\bibitem [{\citenamefont {Komissarov}\ \emph {et~al.}(2024)\citenamefont
  {Komissarov}, \citenamefont {Holder},\ and\ \citenamefont
  {Queiroz}}]{Komissarov24}%
  \BibitemOpen
  \bibfield  {author} {\bibinfo {author} {\bibfnamefont {I.}~\bibnamefont
  {Komissarov}}, \bibinfo {author} {\bibfnamefont {T.}~\bibnamefont {Holder}},\
  and\ \bibinfo {author} {\bibfnamefont {R.}~\bibnamefont {Queiroz}},\
  }\bibfield  {title} {\bibinfo {title} {The quantum geometric origin of
  capacitance in insulators},\ }\href
  {https://doi.org/10.1038/s41467-024-48808-x} {\bibfield  {journal} {\bibinfo
  {journal} {Nat. Commun.}\ }\textbf {\bibinfo {volume} {15}},\ \bibinfo
  {pages} {4621} (\bibinfo {year} {2024})}\BibitemShut {NoStop}%
\bibitem [{\citenamefont {Chen}(2025)}]{Chen25_optical_marker}%
  \BibitemOpen
  \bibfield  {author} {\bibinfo {author} {\bibfnamefont {W.}~\bibnamefont
  {Chen}},\ }\bibfield  {title} {\bibinfo {title} {Dielectric and optical
  markers originating from quantum geometry},\ }\href
  {https://doi.org/10.1103/PhysRevB.111.085202} {\bibfield  {journal} {\bibinfo
   {journal} {Phys. Rev. B}\ }\textbf {\bibinfo {volume} {111}},\ \bibinfo
  {pages} {085202} (\bibinfo {year} {2025})}\BibitemShut {NoStop}%
\bibitem [{\citenamefont {Provost}\ and\ \citenamefont
  {Vallee}(1980)}]{Provost80}%
  \BibitemOpen
  \bibfield  {author} {\bibinfo {author} {\bibfnamefont {J.~P.}\ \bibnamefont
  {Provost}}\ and\ \bibinfo {author} {\bibfnamefont {G.}~\bibnamefont
  {Vallee}},\ }\bibfield  {title} {\bibinfo {title} {Riemannian structure on
  manifolds of quantum states},\ }\href
  {https://projecteuclid.org:443/euclid.cmp/1103908308} {\bibfield  {journal}
  {\bibinfo  {journal} {Comm. Math. Phys.}\ }\textbf {\bibinfo {volume} {76}},\
  \bibinfo {pages} {289} (\bibinfo {year} {1980})}\BibitemShut {NoStop}%
\bibitem [{\citenamefont {Berry}(1984)}]{Berry84}%
  \BibitemOpen
  \bibfield  {author} {\bibinfo {author} {\bibfnamefont {M.~V.}\ \bibnamefont
  {Berry}},\ }\bibfield  {title} {\bibinfo {title} {Quantal phase factors
  accompanying adiabatic changes},\ }\href
  {https://doi.org/10.1098/rspa.1984.0023} {\bibfield  {journal} {\bibinfo
  {journal} {Proc. R. Soc. A}\ }\textbf {\bibinfo {volume} {392}},\ \bibinfo
  {pages} {45} (\bibinfo {year} {1984})}\BibitemShut {NoStop}%
\bibitem [{\citenamefont {Roy}(2014)}]{Roy14}%
  \BibitemOpen
  \bibfield  {author} {\bibinfo {author} {\bibfnamefont {R.}~\bibnamefont
  {Roy}},\ }\bibfield  {title} {\bibinfo {title} {Band geometry of fractional
  topological insulators},\ }\href {https://doi.org/10.1103/PhysRevB.90.165139}
  {\bibfield  {journal} {\bibinfo  {journal} {Phys. Rev. B}\ }\textbf {\bibinfo
  {volume} {90}},\ \bibinfo {pages} {165139} (\bibinfo {year}
  {2014})}\BibitemShut {NoStop}%
\bibitem [{\citenamefont {Peotta}\ and\ \citenamefont
  {T\"{o}rm\"{a}}(2015)}]{Peotta15}%
  \BibitemOpen
  \bibfield  {author} {\bibinfo {author} {\bibfnamefont {S.}~\bibnamefont
  {Peotta}}\ and\ \bibinfo {author} {\bibfnamefont {P.}~\bibnamefont
  {T\"{o}rm\"{a}}},\ }\bibfield  {title} {\bibinfo {title} {Superfluidity in
  topologically nontrivial flat bands},\ }\href
  {https://doi.org/10.1038/ncomms9944} {\bibfield  {journal} {\bibinfo
  {journal} {Nat. Commun.}\ }\textbf {\bibinfo {volume} {6}},\ \bibinfo {pages}
  {8944} (\bibinfo {year} {2015})}\BibitemShut {NoStop}%
\bibitem [{\citenamefont {Ozawa}\ and\ \citenamefont {Mera}(2021)}]{Ozawa21}%
  \BibitemOpen
  \bibfield  {author} {\bibinfo {author} {\bibfnamefont {T.}~\bibnamefont
  {Ozawa}}\ and\ \bibinfo {author} {\bibfnamefont {B.}~\bibnamefont {Mera}},\
  }\bibfield  {title} {\bibinfo {title} {Relations between topology and the
  quantum metric for chern insulators},\ }\href
  {https://doi.org/10.1103/PhysRevB.104.045103} {\bibfield  {journal} {\bibinfo
   {journal} {Phys. Rev. B}\ }\textbf {\bibinfo {volume} {104}},\ \bibinfo
  {pages} {045103} (\bibinfo {year} {2021})}\BibitemShut {NoStop}%
\bibitem [{\citenamefont {{Ma, Yu-Quan}}\ \emph {et~al.}(2013)\citenamefont
  {{Ma, Yu-Quan}}, \citenamefont {{Gu, Shi-Jian}}, \citenamefont {{Chen, Shu}},
  \citenamefont {{Fan, Heng}},\ and\ \citenamefont {{Liu, Wu-Ming}}}]{Ma13}%
  \BibitemOpen
  \bibfield  {author} {\bibinfo {author} {\bibnamefont {{Ma, Yu-Quan}}},
  \bibinfo {author} {\bibnamefont {{Gu, Shi-Jian}}}, \bibinfo {author}
  {\bibnamefont {{Chen, Shu}}}, \bibinfo {author} {\bibnamefont {{Fan,
  Heng}}},\ and\ \bibinfo {author} {\bibnamefont {{Liu, Wu-Ming}}},\ }\bibfield
   {title} {\bibinfo {title} {The euler number of bloch states manifold and the
  quantum phases in gapped fermionic systems},\ }\href
  {https://doi.org/10.1209/0295-5075/103/10008} {\bibfield  {journal} {\bibinfo
   {journal} {EPL}\ }\textbf {\bibinfo {volume} {103}},\ \bibinfo {pages}
  {10008} (\bibinfo {year} {2013})}\BibitemShut {NoStop}%
\bibitem [{\citenamefont {Ma}(2014)}]{Ma14}%
  \BibitemOpen
  \bibfield  {author} {\bibinfo {author} {\bibfnamefont {Y.-Q.}\ \bibnamefont
  {Ma}},\ }\bibfield  {title} {\bibinfo {title} {Quantum distance and the euler
  number index of the bloch band in a one-dimensional spin model},\ }\href
  {https://doi.org/10.1103/PhysRevE.90.042133} {\bibfield  {journal} {\bibinfo
  {journal} {Phys. Rev. E}\ }\textbf {\bibinfo {volume} {90}},\ \bibinfo
  {pages} {042133} (\bibinfo {year} {2014})}\BibitemShut {NoStop}%
\bibitem [{\citenamefont {Yang}\ \emph {et~al.}(2015)\citenamefont {Yang},
  \citenamefont {Ma},\ and\ \citenamefont {Li}}]{Yang15}%
  \BibitemOpen
  \bibfield  {author} {\bibinfo {author} {\bibfnamefont {L.}~\bibnamefont
  {Yang}}, \bibinfo {author} {\bibfnamefont {Y.-Q.}\ \bibnamefont {Ma}},\ and\
  \bibinfo {author} {\bibfnamefont {X.-G.}\ \bibnamefont {Li}},\ }\bibfield
  {title} {\bibinfo {title} {Geometric tensor and the topological
  characterization of the bloch band in a two-band lattice model},\ }\href
  {https://doi.org/https://doi.org/10.1016/j.physb.2014.09.022} {\bibfield
  {journal} {\bibinfo  {journal} {Physica B Condens. Matter}\ }\textbf
  {\bibinfo {volume} {456}},\ \bibinfo {pages} {359} (\bibinfo {year}
  {2015})}\BibitemShut {NoStop}%
\bibitem [{\citenamefont {Pi\'echon}\ \emph {et~al.}(2016)\citenamefont
  {Pi\'echon}, \citenamefont {Raoux}, \citenamefont {Fuchs},\ and\
  \citenamefont {Montambaux}}]{Piechon16}%
  \BibitemOpen
  \bibfield  {author} {\bibinfo {author} {\bibfnamefont {F.}~\bibnamefont
  {Pi\'echon}}, \bibinfo {author} {\bibfnamefont {A.}~\bibnamefont {Raoux}},
  \bibinfo {author} {\bibfnamefont {J.-N.}\ \bibnamefont {Fuchs}},\ and\
  \bibinfo {author} {\bibfnamefont {G.}~\bibnamefont {Montambaux}},\ }\bibfield
   {title} {\bibinfo {title} {Geometric orbital susceptibility: Quantum metric
  without berry curvature},\ }\href
  {https://doi.org/10.1103/PhysRevB.94.134423} {\bibfield  {journal} {\bibinfo
  {journal} {Phys. Rev. B}\ }\textbf {\bibinfo {volume} {94}},\ \bibinfo
  {pages} {134423} (\bibinfo {year} {2016})}\BibitemShut {NoStop}%
\bibitem [{\citenamefont {Palumbo}\ and\ \citenamefont
  {Goldman}(2018)}]{Palumbo18}%
  \BibitemOpen
  \bibfield  {author} {\bibinfo {author} {\bibfnamefont {G.}~\bibnamefont
  {Palumbo}}\ and\ \bibinfo {author} {\bibfnamefont {N.}~\bibnamefont
  {Goldman}},\ }\bibfield  {title} {\bibinfo {title} {Revealing tensor
  monopoles through quantum-metric measurements},\ }\href
  {https://doi.org/10.1103/PhysRevLett.121.170401} {\bibfield  {journal}
  {\bibinfo  {journal} {Phys. Rev. Lett.}\ }\textbf {\bibinfo {volume} {121}},\
  \bibinfo {pages} {170401} (\bibinfo {year} {2018})}\BibitemShut {NoStop}%
\bibitem [{\citenamefont {Panahiyan}\ \emph {et~al.}(2020)\citenamefont
  {Panahiyan}, \citenamefont {Chen},\ and\ \citenamefont
  {Fritzsche}}]{Panahiyan20_fidelity_susceptibility}%
  \BibitemOpen
  \bibfield  {author} {\bibinfo {author} {\bibfnamefont {S.}~\bibnamefont
  {Panahiyan}}, \bibinfo {author} {\bibfnamefont {W.}~\bibnamefont {Chen}},\
  and\ \bibinfo {author} {\bibfnamefont {S.}~\bibnamefont {Fritzsche}},\
  }\bibfield  {title} {\bibinfo {title} {Fidelity susceptibility near
  topological phase transitions in quantum walks},\ }\href
  {https://doi.org/10.1103/PhysRevB.102.134111} {\bibfield  {journal} {\bibinfo
   {journal} {Phys. Rev. B}\ }\textbf {\bibinfo {volume} {102}},\ \bibinfo
  {pages} {134111} (\bibinfo {year} {2020})}\BibitemShut {NoStop}%
\bibitem [{\citenamefont {Ma}(2020)}]{Ma20}%
  \BibitemOpen
  \bibfield  {author} {\bibinfo {author} {\bibfnamefont {Y.-Q.}\ \bibnamefont
  {Ma}},\ }\bibfield  {title} {\bibinfo {title} {Euler characteristic number of
  the energy band and the reason for its non-integer values},\ }\href@noop {}
  {\bibfield  {journal} {\bibinfo  {journal} {arXiv:2001.05946}\ } (\bibinfo
  {year} {2020})}\BibitemShut {NoStop}%
\bibitem [{\citenamefont {Salerno}\ \emph {et~al.}(2020)\citenamefont
  {Salerno}, \citenamefont {Goldman},\ and\ \citenamefont
  {Palumbo}}]{Salerno20}%
  \BibitemOpen
  \bibfield  {author} {\bibinfo {author} {\bibfnamefont {G.}~\bibnamefont
  {Salerno}}, \bibinfo {author} {\bibfnamefont {N.}~\bibnamefont {Goldman}},\
  and\ \bibinfo {author} {\bibfnamefont {G.}~\bibnamefont {Palumbo}},\
  }\bibfield  {title} {\bibinfo {title} {Floquet-engineering of nodal rings and
  nodal spheres and their characterization using the quantum metric},\ }\href
  {https://doi.org/10.1103/PhysRevResearch.2.013224} {\bibfield  {journal}
  {\bibinfo  {journal} {Phys. Rev. Research}\ }\textbf {\bibinfo {volume}
  {2}},\ \bibinfo {pages} {013224} (\bibinfo {year} {2020})}\BibitemShut
  {NoStop}%
\bibitem [{\citenamefont {Mera}\ \emph {et~al.}(2022)\citenamefont {Mera},
  \citenamefont {Zhang},\ and\ \citenamefont {Goldman}}]{Mera22}%
  \BibitemOpen
  \bibfield  {author} {\bibinfo {author} {\bibfnamefont {B.}~\bibnamefont
  {Mera}}, \bibinfo {author} {\bibfnamefont {A.}~\bibnamefont {Zhang}},\ and\
  \bibinfo {author} {\bibfnamefont {N.}~\bibnamefont {Goldman}},\ }\bibfield
  {title} {\bibinfo {title} {Relating the topology of dirac hamiltonians to
  quantum geometry: When the quantum metric dictates chern numbers and winding
  numbers},\ }\href {https://doi.org/10.21468/SciPostPhys.12.1.018} {\bibfield
  {journal} {\bibinfo  {journal} {SciPost Phys.}\ }\textbf {\bibinfo {volume}
  {12}},\ \bibinfo {pages} {018} (\bibinfo {year} {2022})}\BibitemShut
  {NoStop}%
\bibitem [{\citenamefont {von Gersdorff}\ and\ \citenamefont
  {Chen}(2021)}]{vonGersdorff21_metric_curvature}%
  \BibitemOpen
  \bibfield  {author} {\bibinfo {author} {\bibfnamefont {G.}~\bibnamefont {von
  Gersdorff}}\ and\ \bibinfo {author} {\bibfnamefont {W.}~\bibnamefont
  {Chen}},\ }\bibfield  {title} {\bibinfo {title} {Measurement of topological
  order based on metric-curvature correspondence},\ }\href
  {https://doi.org/10.1103/PhysRevB.104.195133} {\bibfield  {journal} {\bibinfo
   {journal} {Phys. Rev. B}\ }\textbf {\bibinfo {volume} {104}},\ \bibinfo
  {pages} {195133} (\bibinfo {year} {2021})}\BibitemShut {NoStop}%
\bibitem [{\citenamefont {Chen}(2026)}]{Chen26_QCRB}%
  \BibitemOpen
  \bibfield  {author} {\bibinfo {author} {\bibfnamefont {W.}~\bibnamefont
  {Chen}},\ }\bibfield  {title} {\bibinfo {title} {Quantum cram\'er-rao bound
  on quantum metric as a single-observable uncertainty relation},\ }\href
  {https://doi.org/10.1103/kknq-k13g} {\bibfield  {journal} {\bibinfo
  {journal} {Phys. Rev. B}\ }\textbf {\bibinfo {volume} {114}},\ \bibinfo
  {pages} {235118} (\bibinfo {year} {2026})}\BibitemShut {NoStop}%
\bibitem [{\citenamefont {Helstrom}(1967)}]{Helstrom67}%
  \BibitemOpen
  \bibfield  {author} {\bibinfo {author} {\bibfnamefont {C.}~\bibnamefont
  {Helstrom}},\ }\bibfield  {title} {\bibinfo {title} {Minimum mean-squared
  error of estimates in quantum statistics},\ }\href
  {https://doi.org/https://doi.org/10.1016/0375-9601(67)90366-0} {\bibfield
  {journal} {\bibinfo  {journal} {Physics Letters A}\ }\textbf {\bibinfo
  {volume} {25}},\ \bibinfo {pages} {101} (\bibinfo {year} {1967})}\BibitemShut
  {NoStop}%
\bibitem [{\citenamefont {Hovelo}(2011)}]{Hovelo11}%
  \BibitemOpen
  \bibfield  {author} {\bibinfo {author} {\bibfnamefont {A.}~\bibnamefont
  {Hovelo}},\ }\href@noop {} {\emph {\bibinfo {title} {Probabilistic and
  Statistical Aspects of Quantum Theory}}}\ (\bibinfo  {publisher} {Edizioni
  della Normale Pisa},\ \bibinfo {year} {2011})\BibitemShut {NoStop}%
\bibitem [{\citenamefont {Braunstein}\ and\ \citenamefont
  {Caves}(1994)}]{Braunstein94}%
  \BibitemOpen
  \bibfield  {author} {\bibinfo {author} {\bibfnamefont {S.~L.}\ \bibnamefont
  {Braunstein}}\ and\ \bibinfo {author} {\bibfnamefont {C.~M.}\ \bibnamefont
  {Caves}},\ }\bibfield  {title} {\bibinfo {title} {Statistical distance and
  the geometry of quantum states},\ }\href
  {https://doi.org/10.1103/PhysRevLett.72.3439} {\bibfield  {journal} {\bibinfo
   {journal} {Phys. Rev. Lett.}\ }\textbf {\bibinfo {volume} {72}},\ \bibinfo
  {pages} {3439} (\bibinfo {year} {1994})}\BibitemShut {NoStop}%
\bibitem [{\citenamefont {Braunstein}\ \emph {et~al.}(1996)\citenamefont
  {Braunstein}, \citenamefont {Caves},\ and\ \citenamefont
  {Milburn}}]{Braunstein96}%
  \BibitemOpen
  \bibfield  {author} {\bibinfo {author} {\bibfnamefont {S.~L.}\ \bibnamefont
  {Braunstein}}, \bibinfo {author} {\bibfnamefont {C.~M.}\ \bibnamefont
  {Caves}},\ and\ \bibinfo {author} {\bibfnamefont {G.}~\bibnamefont
  {Milburn}},\ }\bibfield  {title} {\bibinfo {title} {Generalized uncertainty
  relations: Theory, examples, and lorentz invariance},\ }\href
  {https://doi.org/https://doi.org/10.1006/aphy.1996.0040} {\bibfield
  {journal} {\bibinfo  {journal} {Annals of Physics}\ }\textbf {\bibinfo
  {volume} {247}},\ \bibinfo {pages} {135} (\bibinfo {year}
  {1996})}\BibitemShut {NoStop}%
\bibitem [{\citenamefont {Paris}(2009)}]{Paris09}%
  \BibitemOpen
  \bibfield  {author} {\bibinfo {author} {\bibfnamefont {M.~G.~A.}\
  \bibnamefont {Paris}},\ }\bibfield  {title} {\bibinfo {title} {Quantum
  estimation for quantum technology},\ }\href
  {https://doi.org/10.1142/S0219749909004839} {\bibfield  {journal} {\bibinfo
  {journal} {Int. J. Quantum Inf.}\ }\textbf {\bibinfo {volume} {07}},\
  \bibinfo {pages} {125} (\bibinfo {year} {2009})}\BibitemShut {NoStop}%
\bibitem [{\citenamefont {Liu}\ \emph {et~al.}(2019)\citenamefont {Liu},
  \citenamefont {Yuan}, \citenamefont {Lu},\ and\ \citenamefont
  {Wang}}]{Liu20}%
  \BibitemOpen
  \bibfield  {author} {\bibinfo {author} {\bibfnamefont {J.}~\bibnamefont
  {Liu}}, \bibinfo {author} {\bibfnamefont {H.}~\bibnamefont {Yuan}}, \bibinfo
  {author} {\bibfnamefont {X.-M.}\ \bibnamefont {Lu}},\ and\ \bibinfo {author}
  {\bibfnamefont {X.}~\bibnamefont {Wang}},\ }\bibfield  {title} {\bibinfo
  {title} {Quantum fisher information matrix and multiparameter estimation},\
  }\href {https://doi.org/10.1088/1751-8121/ab5d4d} {\bibfield  {journal}
  {\bibinfo  {journal} {J. Phys. A Math. Theor.}\ }\textbf {\bibinfo {volume}
  {53}},\ \bibinfo {pages} {023001} (\bibinfo {year} {2019})}\BibitemShut
  {NoStop}%
\bibitem [{\citenamefont {Uhlmann}(1986)}]{Uhlmann86}%
  \BibitemOpen
  \bibfield  {author} {\bibinfo {author} {\bibfnamefont {A.}~\bibnamefont
  {Uhlmann}},\ }\bibfield  {title} {\bibinfo {title} {Parallel transport and
  “quantum holonomy” along density operators},\ }\href
  {https://doi.org/https://doi.org/10.1016/0034-4877(86)90055-8} {\bibfield
  {journal} {\bibinfo  {journal} {Rep. Math. Phys.}\ }\textbf {\bibinfo
  {volume} {24}},\ \bibinfo {pages} {229} (\bibinfo {year} {1986})}\BibitemShut
  {NoStop}%
\bibitem [{\citenamefont {Uhlmann}(1991)}]{Uhlmann91}%
  \BibitemOpen
  \bibfield  {author} {\bibinfo {author} {\bibfnamefont {A.}~\bibnamefont
  {Uhlmann}},\ }\bibfield  {title} {\bibinfo {title} {A gauge field governing
  parallel transport along mixed states},\ }\href
  {https://doi.org/10.1007/BF00420373} {\bibfield  {journal} {\bibinfo
  {journal} {Lett. Math. Phys.}\ }\textbf {\bibinfo {volume} {21}},\ \bibinfo
  {pages} {229} (\bibinfo {year} {1991})}\BibitemShut {NoStop}%
\bibitem [{\citenamefont {Carollo}\ \emph {et~al.}(2018)\citenamefont
  {Carollo}, \citenamefont {Spagnolo},\ and\ \citenamefont
  {Valenti}}]{Carollo18}%
  \BibitemOpen
  \bibfield  {author} {\bibinfo {author} {\bibfnamefont {A.}~\bibnamefont
  {Carollo}}, \bibinfo {author} {\bibfnamefont {B.}~\bibnamefont {Spagnolo}},\
  and\ \bibinfo {author} {\bibfnamefont {D.}~\bibnamefont {Valenti}},\
  }\bibfield  {title} {\bibinfo {title} {Uhlmann curvature in dissipative phase
  transitions},\ }\href {https://doi.org/10.1038/s41598-018-27362-9} {\bibfield
   {journal} {\bibinfo  {journal} {Sci. Rep.}\ }\textbf {\bibinfo {volume}
  {8}},\ \bibinfo {pages} {9852} (\bibinfo {year} {2018})}\BibitemShut
  {NoStop}%
\bibitem [{\citenamefont {Ercolessi}\ and\ \citenamefont
  {Schiavina}(2012)}]{Ercolessi12}%
  \BibitemOpen
  \bibfield  {author} {\bibinfo {author} {\bibfnamefont {E.}~\bibnamefont
  {Ercolessi}}\ and\ \bibinfo {author} {\bibfnamefont {M.}~\bibnamefont
  {Schiavina}},\ }\bibfield  {title} {\bibinfo {title} {Geometry of mixed
  states for a q-bit and the quantum fisher information tensor},\ }\href
  {https://doi.org/10.1088/1751-8113/45/36/365303} {\bibfield  {journal}
  {\bibinfo  {journal} {J. Phys. A: Math. Theor.}\ }\textbf {\bibinfo {volume}
  {45}},\ \bibinfo {pages} {365303} (\bibinfo {year} {2012})}\BibitemShut
  {NoStop}%
\bibitem [{\citenamefont {Het\'enyi}\ and\ \citenamefont
  {L\'evay}(2023)}]{Hetenyi23}%
  \BibitemOpen
  \bibfield  {author} {\bibinfo {author} {\bibfnamefont {B.}~\bibnamefont
  {Het\'enyi}}\ and\ \bibinfo {author} {\bibfnamefont {P.}~\bibnamefont
  {L\'evay}},\ }\bibfield  {title} {\bibinfo {title} {Fluctuations, uncertainty
  relations, and the geometry of quantum state manifolds},\ }\href
  {https://doi.org/10.1103/PhysRevA.108.032218} {\bibfield  {journal} {\bibinfo
   {journal} {Phys. Rev. A}\ }\textbf {\bibinfo {volume} {108}},\ \bibinfo
  {pages} {032218} (\bibinfo {year} {2023})}\BibitemShut {NoStop}%
\bibitem [{\citenamefont {Het\'enyi}(2026)}]{Hetenyi26}%
  \BibitemOpen
  \bibfield  {author} {\bibinfo {author} {\bibfnamefont {B.}~\bibnamefont
  {Het\'enyi}},\ }\bibfield  {title} {\bibinfo {title} {From generating
  functions to the geometric binder cumulant},\ }\href
  {https://doi.org/10.1088/1361-648X/ae5c50} {\bibfield  {journal} {\bibinfo
  {journal} {J. Phys. Condens. Matter}\ }\textbf {\bibinfo {volume} {38}},\
  \bibinfo {pages} {153002} (\bibinfo {year} {2026})}\BibitemShut {NoStop}%
\bibitem [{\citenamefont {Abreu}\ and\ \citenamefont
  {Chen}(2026)}]{Abreu26_generating_fn}%
  \BibitemOpen
  \bibfield  {author} {\bibinfo {author} {\bibfnamefont {F.~P.}\ \bibnamefont
  {Abreu}}\ and\ \bibinfo {author} {\bibfnamefont {W.}~\bibnamefont {Chen}},\
  }\href {https://arxiv.org/abs/2511.05260} {\bibinfo {title} {Generating
  function and bloch representation for quantum fisher tensor}} (\bibinfo
  {year} {2026}),\ \Eprint {https://arxiv.org/abs/2511.05260} {arXiv:2511.05260
  [quant-ph]} \BibitemShut {NoStop}%
\bibitem [{\citenamefont {Robertson}(1929)}]{Robertson29}%
  \BibitemOpen
  \bibfield  {author} {\bibinfo {author} {\bibfnamefont {H.~P.}\ \bibnamefont
  {Robertson}},\ }\bibfield  {title} {\bibinfo {title} {The uncertainty
  principle},\ }\href {https://doi.org/10.1103/PhysRev.34.163} {\bibfield
  {journal} {\bibinfo  {journal} {Phys. Rev.}\ }\textbf {\bibinfo {volume}
  {34}},\ \bibinfo {pages} {163} (\bibinfo {year} {1929})}\BibitemShut
  {NoStop}%
\bibitem [{\citenamefont {Schr{\"o}dinger}(1930)}]{Schrodinger30}%
  \BibitemOpen
  \bibfield  {author} {\bibinfo {author} {\bibfnamefont {E.}~\bibnamefont
  {Schr{\"o}dinger}},\ }\bibfield  {title} {\bibinfo {title} {Zum
  heisenbergschen unsch{\"a}rfeprinzip},\ }\href@noop {} {\bibfield  {journal}
  {\bibinfo  {journal} {Sitzungsber. Preuss. Akad. Wiss., Phys.-Math. Kl.}\
  }\textbf {\bibinfo {volume} {14}},\ \bibinfo {pages} {296} (\bibinfo {year}
  {1930})}\BibitemShut {NoStop}%
\bibitem [{\citenamefont {Efimov}(1976)}]{Efimov76}%
  \BibitemOpen
  \bibfield  {author} {\bibinfo {author} {\bibfnamefont {S.~P.}\ \bibnamefont
  {Efimov}},\ }\bibfield  {title} {\bibinfo {title} {Mathematical formulation
  of indeterminacy relations},\ }\href {https://doi.org/10.1007/BF00945688}
  {\bibfield  {journal} {\bibinfo  {journal} {Sov. Phys. J.}\ }\textbf
  {\bibinfo {volume} {19}},\ \bibinfo {pages} {340} (\bibinfo {year}
  {1976})}\BibitemShut {NoStop}%
\bibitem [{\citenamefont {Dodonov}(2018)}]{Dodonov18}%
  \BibitemOpen
  \bibfield  {author} {\bibinfo {author} {\bibfnamefont {V.~V.}\ \bibnamefont
  {Dodonov}},\ }\bibfield  {title} {\bibinfo {title} {Variance uncertainty
  relations without covariances for three and four observables},\ }\href
  {https://doi.org/10.1103/PhysRevA.97.022105} {\bibfield  {journal} {\bibinfo
  {journal} {Phys. Rev. A}\ }\textbf {\bibinfo {volume} {97}},\ \bibinfo
  {pages} {022105} (\bibinfo {year} {2018})}\BibitemShut {NoStop}%
\bibitem [{\citenamefont {Liu}\ \emph {et~al.}(2013)\citenamefont {Liu},
  \citenamefont {Shan}, \citenamefont {Yao}, \citenamefont {Yao},\ and\
  \citenamefont {Xiao}}]{Liu13}%
  \BibitemOpen
  \bibfield  {author} {\bibinfo {author} {\bibfnamefont {G.-B.}\ \bibnamefont
  {Liu}}, \bibinfo {author} {\bibfnamefont {W.-Y.}\ \bibnamefont {Shan}},
  \bibinfo {author} {\bibfnamefont {Y.}~\bibnamefont {Yao}}, \bibinfo {author}
  {\bibfnamefont {W.}~\bibnamefont {Yao}},\ and\ \bibinfo {author}
  {\bibfnamefont {D.}~\bibnamefont {Xiao}},\ }\bibfield  {title} {\bibinfo
  {title} {Three-band tight-binding model for monolayers of group-vib
  transition metal dichalcogenides},\ }\href
  {https://doi.org/10.1103/PhysRevB.88.085433} {\bibfield  {journal} {\bibinfo
  {journal} {Phys. Rev. B}\ }\textbf {\bibinfo {volume} {88}},\ \bibinfo
  {pages} {085433} (\bibinfo {year} {2013})}\BibitemShut {NoStop}%
\bibitem [{\citenamefont {Vogl}\ \emph {et~al.}(1983)\citenamefont {Vogl},
  \citenamefont {Hjalmarson},\ and\ \citenamefont {Dow}}]{Vogl83}%
  \BibitemOpen
  \bibfield  {author} {\bibinfo {author} {\bibfnamefont {P.}~\bibnamefont
  {Vogl}}, \bibinfo {author} {\bibfnamefont {H.~P.}\ \bibnamefont
  {Hjalmarson}},\ and\ \bibinfo {author} {\bibfnamefont {J.~D.}\ \bibnamefont
  {Dow}},\ }\bibfield  {title} {\bibinfo {title} {A semi-empirical
  tight-binding theory of the electronic structure of semiconductors†},\
  }\href {https://doi.org/https://doi.org/10.1016/0022-3697(83)90064-1}
  {\bibfield  {journal} {\bibinfo  {journal} {J. Phys. Chem. Sol.}\ }\textbf
  {\bibinfo {volume} {44}},\ \bibinfo {pages} {365} (\bibinfo {year}
  {1983})}\BibitemShut {NoStop}%
\bibitem [{\citenamefont {Graf}\ and\ \citenamefont {Vogl}(1995)}]{Graf95}%
  \BibitemOpen
  \bibfield  {author} {\bibinfo {author} {\bibfnamefont {M.}~\bibnamefont
  {Graf}}\ and\ \bibinfo {author} {\bibfnamefont {P.}~\bibnamefont {Vogl}},\
  }\bibfield  {title} {\bibinfo {title} {Electromagnetic fields and dielectric
  response in empirical tight-binding theory},\ }\href
  {https://doi.org/10.1103/PhysRevB.51.4940} {\bibfield  {journal} {\bibinfo
  {journal} {Phys. Rev. B}\ }\textbf {\bibinfo {volume} {51}},\ \bibinfo
  {pages} {4940} (\bibinfo {year} {1995})}\BibitemShut {NoStop}%
\bibitem [{\citenamefont {C\'ardenas-Castillo}\ \emph
  {et~al.}(2024)\citenamefont {C\'ardenas-Castillo}, \citenamefont {Zhang},
  \citenamefont {Freire}, \citenamefont {Kochan},\ and\ \citenamefont
  {Chen}}]{CardenasCastillo24_spread_Wannier}%
  \BibitemOpen
  \bibfield  {author} {\bibinfo {author} {\bibfnamefont {L.~F.}\ \bibnamefont
  {C\'ardenas-Castillo}}, \bibinfo {author} {\bibfnamefont {S.}~\bibnamefont
  {Zhang}}, \bibinfo {author} {\bibfnamefont {F.~L.}\ \bibnamefont {Freire}},
  \bibinfo {author} {\bibfnamefont {D.}~\bibnamefont {Kochan}},\ and\ \bibinfo
  {author} {\bibfnamefont {W.}~\bibnamefont {Chen}},\ }\bibfield  {title}
  {\bibinfo {title} {Detecting the spread of valence-band wannier functions by
  optical sum rules},\ }\href {https://doi.org/10.1103/PhysRevB.110.075203}
  {\bibfield  {journal} {\bibinfo  {journal} {Phys. Rev. B}\ }\textbf {\bibinfo
  {volume} {110}},\ \bibinfo {pages} {075203} (\bibinfo {year}
  {2024})}\BibitemShut {NoStop}%
\bibitem [{\citenamefont {Marzari}\ and\ \citenamefont
  {Vanderbilt}(1997)}]{Marzari97}%
  \BibitemOpen
  \bibfield  {author} {\bibinfo {author} {\bibfnamefont {N.}~\bibnamefont
  {Marzari}}\ and\ \bibinfo {author} {\bibfnamefont {D.}~\bibnamefont
  {Vanderbilt}},\ }\bibfield  {title} {\bibinfo {title} {Maximally localized
  generalized wannier functions for composite energy bands},\ }\href
  {https://doi.org/10.1103/PhysRevB.56.12847} {\bibfield  {journal} {\bibinfo
  {journal} {Phys. Rev. B}\ }\textbf {\bibinfo {volume} {56}},\ \bibinfo
  {pages} {12847} (\bibinfo {year} {1997})}\BibitemShut {NoStop}%
\bibitem [{\citenamefont {Souza}\ and\ \citenamefont
  {Vanderbilt}(2008)}]{Souza08}%
  \BibitemOpen
  \bibfield  {author} {\bibinfo {author} {\bibfnamefont {I.}~\bibnamefont
  {Souza}}\ and\ \bibinfo {author} {\bibfnamefont {D.}~\bibnamefont
  {Vanderbilt}},\ }\bibfield  {title} {\bibinfo {title} {Dichroic $f$-sum rule
  and the orbital magnetization of crystals},\ }\href
  {https://doi.org/10.1103/PhysRevB.77.054438} {\bibfield  {journal} {\bibinfo
  {journal} {Phys. Rev. B}\ }\textbf {\bibinfo {volume} {77}},\ \bibinfo
  {pages} {054438} (\bibinfo {year} {2008})}\BibitemShut {NoStop}%
\bibitem [{\citenamefont {Marzari}\ \emph {et~al.}(2012)\citenamefont
  {Marzari}, \citenamefont {Mostofi}, \citenamefont {Yates}, \citenamefont
  {Souza},\ and\ \citenamefont {Vanderbilt}}]{Marzari12}%
  \BibitemOpen
  \bibfield  {author} {\bibinfo {author} {\bibfnamefont {N.}~\bibnamefont
  {Marzari}}, \bibinfo {author} {\bibfnamefont {A.~A.}\ \bibnamefont
  {Mostofi}}, \bibinfo {author} {\bibfnamefont {J.~R.}\ \bibnamefont {Yates}},
  \bibinfo {author} {\bibfnamefont {I.}~\bibnamefont {Souza}},\ and\ \bibinfo
  {author} {\bibfnamefont {D.}~\bibnamefont {Vanderbilt}},\ }\bibfield  {title}
  {\bibinfo {title} {Maximally localized wannier functions: Theory and
  applications},\ }\href {https://doi.org/10.1103/RevModPhys.84.1419}
  {\bibfield  {journal} {\bibinfo  {journal} {Rev. Mod. Phys.}\ }\textbf
  {\bibinfo {volume} {84}},\ \bibinfo {pages} {1419} (\bibinfo {year}
  {2012})}\BibitemShut {NoStop}%
\bibitem [{\citenamefont {Watanabe}\ \emph {et~al.}(2010)\citenamefont
  {Watanabe}, \citenamefont {Sagawa},\ and\ \citenamefont {Ueda}}]{Watanabe10}%
  \BibitemOpen
  \bibfield  {author} {\bibinfo {author} {\bibfnamefont {Y.}~\bibnamefont
  {Watanabe}}, \bibinfo {author} {\bibfnamefont {T.}~\bibnamefont {Sagawa}},\
  and\ \bibinfo {author} {\bibfnamefont {M.}~\bibnamefont {Ueda}},\ }\bibfield
  {title} {\bibinfo {title} {Optimal measurement on noisy quantum systems},\
  }\href {https://doi.org/10.1103/PhysRevLett.104.020401} {\bibfield  {journal}
  {\bibinfo  {journal} {Phys. Rev. Lett.}\ }\textbf {\bibinfo {volume} {104}},\
  \bibinfo {pages} {020401} (\bibinfo {year} {2010})}\BibitemShut {NoStop}%
\bibitem [{\citenamefont {Watanabe}\ \emph {et~al.}(2011)\citenamefont
  {Watanabe}, \citenamefont {Sagawa},\ and\ \citenamefont {Ueda}}]{Watanabe11}%
  \BibitemOpen
  \bibfield  {author} {\bibinfo {author} {\bibfnamefont {Y.}~\bibnamefont
  {Watanabe}}, \bibinfo {author} {\bibfnamefont {T.}~\bibnamefont {Sagawa}},\
  and\ \bibinfo {author} {\bibfnamefont {M.}~\bibnamefont {Ueda}},\ }\bibfield
  {title} {\bibinfo {title} {Uncertainty relation revisited from quantum
  estimation theory},\ }\href {https://doi.org/10.1103/PhysRevA.84.042121}
  {\bibfield  {journal} {\bibinfo  {journal} {Phys. Rev. A}\ }\textbf {\bibinfo
  {volume} {84}},\ \bibinfo {pages} {042121} (\bibinfo {year}
  {2011})}\BibitemShut {NoStop}%
\bibitem [{\citenamefont {Ragy}\ \emph {et~al.}(2016)\citenamefont {Ragy},
  \citenamefont {Jarzyna},\ and\ \citenamefont
  {Demkowicz-Dobrza\ifmmode~\acute{n}\else \'{n}\fi{}ski}}]{Ragy16}%
  \BibitemOpen
  \bibfield  {author} {\bibinfo {author} {\bibfnamefont {S.}~\bibnamefont
  {Ragy}}, \bibinfo {author} {\bibfnamefont {M.}~\bibnamefont {Jarzyna}},\ and\
  \bibinfo {author} {\bibfnamefont {R.}~\bibnamefont
  {Demkowicz-Dobrza\ifmmode~\acute{n}\else \'{n}\fi{}ski}},\ }\bibfield
  {title} {\bibinfo {title} {Compatibility in multiparameter quantum
  metrology},\ }\bibfield  {journal} {\bibinfo  {journal} {Phys. Rev. A}\
  }\textbf {\bibinfo {volume} {94}},\ \href
  {https://doi.org/10.1103/PhysRevA.94.052108} {10.1103/PhysRevA.94.052108}
  (\bibinfo {year} {2016})\BibitemShut {NoStop}%
\bibitem [{\citenamefont {Matsuura}\ and\ \citenamefont
  {Ryu}(2010)}]{Matsuura10}%
  \BibitemOpen
  \bibfield  {author} {\bibinfo {author} {\bibfnamefont {S.}~\bibnamefont
  {Matsuura}}\ and\ \bibinfo {author} {\bibfnamefont {S.}~\bibnamefont {Ryu}},\
  }\bibfield  {title} {\bibinfo {title} {Momentum space metric, nonlocal
  operator, and topological insulators},\ }\href
  {https://doi.org/10.1103/PhysRevB.82.245113} {\bibfield  {journal} {\bibinfo
  {journal} {Phys. Rev. B}\ }\textbf {\bibinfo {volume} {82}},\ \bibinfo
  {pages} {245113} (\bibinfo {year} {2010})}\BibitemShut {NoStop}%
\bibitem [{\citenamefont {Porlles}\ and\ \citenamefont
  {Chen}(2023)}]{Porlles23_singlet_SC_metric}%
  \BibitemOpen
  \bibfield  {author} {\bibinfo {author} {\bibfnamefont {D.}~\bibnamefont
  {Porlles}}\ and\ \bibinfo {author} {\bibfnamefont {W.}~\bibnamefont {Chen}},\
  }\bibfield  {title} {\bibinfo {title} {Quantum geometry of singlet
  superconductors},\ }\href {https://doi.org/10.1103/PhysRevB.108.094508}
  {\bibfield  {journal} {\bibinfo  {journal} {Phys. Rev. B}\ }\textbf {\bibinfo
  {volume} {108}},\ \bibinfo {pages} {094508} (\bibinfo {year}
  {2023})}\BibitemShut {NoStop}%
\bibitem [{\citenamefont {de~Sousa}\ \emph {et~al.}(2023)\citenamefont
  {de~Sousa}, \citenamefont {Cruz},\ and\ \citenamefont
  {Chen}}]{deSousa23_fidelity_marker}%
  \BibitemOpen
  \bibfield  {author} {\bibinfo {author} {\bibfnamefont {M.~S.~M.}\
  \bibnamefont {de~Sousa}}, \bibinfo {author} {\bibfnamefont {A.~L.}\
  \bibnamefont {Cruz}},\ and\ \bibinfo {author} {\bibfnamefont
  {W.}~\bibnamefont {Chen}},\ }\bibfield  {title} {\bibinfo {title} {Mapping
  quantum geometry and quantum phase transitions to real space by a fidelity
  marker},\ }\href {https://doi.org/10.1103/PhysRevB.107.205133} {\bibfield
  {journal} {\bibinfo  {journal} {Phys. Rev. B}\ }\textbf {\bibinfo {volume}
  {107}},\ \bibinfo {pages} {205133} (\bibinfo {year} {2023})}\BibitemShut
  {NoStop}%
\bibitem [{\citenamefont {Weng}\ \emph {et~al.}(2015)\citenamefont {Weng},
  \citenamefont {Fang}, \citenamefont {Fang}, \citenamefont {Bernevig},\ and\
  \citenamefont {Dai}}]{Weng15}%
  \BibitemOpen
  \bibfield  {author} {\bibinfo {author} {\bibfnamefont {H.}~\bibnamefont
  {Weng}}, \bibinfo {author} {\bibfnamefont {C.}~\bibnamefont {Fang}}, \bibinfo
  {author} {\bibfnamefont {Z.}~\bibnamefont {Fang}}, \bibinfo {author}
  {\bibfnamefont {B.~A.}\ \bibnamefont {Bernevig}},\ and\ \bibinfo {author}
  {\bibfnamefont {X.}~\bibnamefont {Dai}},\ }\bibfield  {title} {\bibinfo
  {title} {Weyl semimetal phase in noncentrosymmetric transition-metal
  monophosphides},\ }\href {https://doi.org/10.1103/PhysRevX.5.011029}
  {\bibfield  {journal} {\bibinfo  {journal} {Phys. Rev. X}\ }\textbf {\bibinfo
  {volume} {5}},\ \bibinfo {pages} {011029} (\bibinfo {year}
  {2015})}\BibitemShut {NoStop}%
\bibitem [{\citenamefont {Huang}\ \emph {et~al.}(2015)\citenamefont {Huang},
  \citenamefont {Xu}, \citenamefont {Belopolski}, \citenamefont {Lee},
  \citenamefont {Chang}, \citenamefont {Wang}, \citenamefont {Alidoust},
  \citenamefont {Bian}, \citenamefont {Neupane}, \citenamefont {Zhang},
  \citenamefont {Jia}, \citenamefont {Bansil}, \citenamefont {Lin},\ and\
  \citenamefont {Hasan}}]{Huang15_2}%
  \BibitemOpen
  \bibfield  {author} {\bibinfo {author} {\bibfnamefont {S.-M.}\ \bibnamefont
  {Huang}}, \bibinfo {author} {\bibfnamefont {S.-Y.}\ \bibnamefont {Xu}},
  \bibinfo {author} {\bibfnamefont {I.}~\bibnamefont {Belopolski}}, \bibinfo
  {author} {\bibfnamefont {C.-C.}\ \bibnamefont {Lee}}, \bibinfo {author}
  {\bibfnamefont {G.}~\bibnamefont {Chang}}, \bibinfo {author} {\bibfnamefont
  {B.}~\bibnamefont {Wang}}, \bibinfo {author} {\bibfnamefont {N.}~\bibnamefont
  {Alidoust}}, \bibinfo {author} {\bibfnamefont {G.}~\bibnamefont {Bian}},
  \bibinfo {author} {\bibfnamefont {M.}~\bibnamefont {Neupane}}, \bibinfo
  {author} {\bibfnamefont {C.}~\bibnamefont {Zhang}}, \bibinfo {author}
  {\bibfnamefont {S.}~\bibnamefont {Jia}}, \bibinfo {author} {\bibfnamefont
  {A.}~\bibnamefont {Bansil}}, \bibinfo {author} {\bibfnamefont
  {H.}~\bibnamefont {Lin}},\ and\ \bibinfo {author} {\bibfnamefont {M.~Z.}\
  \bibnamefont {Hasan}},\ }\bibfield  {title} {\bibinfo {title} {A weyl fermion
  semimetal with surface fermi arcs in the transition metal monopnictide taas
  class},\ }\href {https://doi.org/10.1038/ncomms8373} {\bibfield  {journal}
  {\bibinfo  {journal} {Nat. Commun.}\ }\textbf {\bibinfo {volume} {6}},\
  \bibinfo {pages} {7373} (\bibinfo {year} {2015})}\BibitemShut {NoStop}%
\bibitem [{\citenamefont {Yan}\ and\ \citenamefont {Felser}(2017)}]{Yan17}%
  \BibitemOpen
  \bibfield  {author} {\bibinfo {author} {\bibfnamefont {B.}~\bibnamefont
  {Yan}}\ and\ \bibinfo {author} {\bibfnamefont {C.}~\bibnamefont {Felser}},\
  }\bibfield  {title} {\bibinfo {title} {Topological materials: Weyl
  semimetals},\ }\href
  {https://doi.org/10.1146/annurev-conmatphys-031016-025458} {\bibfield
  {journal} {\bibinfo  {journal} {Annu. Rev. Condens. Matter Phys.}\ }\textbf
  {\bibinfo {volume} {8}},\ \bibinfo {pages} {337} (\bibinfo {year}
  {2017})}\BibitemShut {NoStop}%
\bibitem [{\citenamefont {Armitage}\ \emph {et~al.}(2018)\citenamefont
  {Armitage}, \citenamefont {Mele},\ and\ \citenamefont
  {Vishwanath}}]{Armitage18}%
  \BibitemOpen
  \bibfield  {author} {\bibinfo {author} {\bibfnamefont {N.~P.}\ \bibnamefont
  {Armitage}}, \bibinfo {author} {\bibfnamefont {E.~J.}\ \bibnamefont {Mele}},\
  and\ \bibinfo {author} {\bibfnamefont {A.}~\bibnamefont {Vishwanath}},\
  }\bibfield  {title} {\bibinfo {title} {Weyl and dirac semimetals in
  three-dimensional solids},\ }\href
  {https://doi.org/10.1103/RevModPhys.90.015001} {\bibfield  {journal}
  {\bibinfo  {journal} {Rev. Mod. Phys.}\ }\textbf {\bibinfo {volume} {90}},\
  \bibinfo {pages} {015001} (\bibinfo {year} {2018})}\BibitemShut {NoStop}%
\bibitem [{\citenamefont {Lv}\ \emph {et~al.}(2021)\citenamefont {Lv},
  \citenamefont {Qian},\ and\ \citenamefont {Ding}}]{Lv21}%
  \BibitemOpen
  \bibfield  {author} {\bibinfo {author} {\bibfnamefont {B.~Q.}\ \bibnamefont
  {Lv}}, \bibinfo {author} {\bibfnamefont {T.}~\bibnamefont {Qian}},\ and\
  \bibinfo {author} {\bibfnamefont {H.}~\bibnamefont {Ding}},\ }\bibfield
  {title} {\bibinfo {title} {Experimental perspective on three-dimensional
  topological semimetals},\ }\href
  {https://doi.org/10.1103/RevModPhys.93.025002} {\bibfield  {journal}
  {\bibinfo  {journal} {Rev. Mod. Phys.}\ }\textbf {\bibinfo {volume} {93}},\
  \bibinfo {pages} {025002} (\bibinfo {year} {2021})}\BibitemShut {NoStop}%
\bibitem [{\citenamefont {Porlles}\ and\ \citenamefont
  {Chen}(2026)}]{Porlles26_semiconductor_metric}%
  \BibitemOpen
  \bibfield  {author} {\bibinfo {author} {\bibfnamefont {D.}~\bibnamefont
  {Porlles}}\ and\ \bibinfo {author} {\bibfnamefont {W.}~\bibnamefont {Chen}},\
  }\bibfield  {title} {\bibinfo {title} {Quantum geometry of common
  semiconductors},\ }\href {https://doi.org/10.1063/5.0307679} {\bibfield
  {journal} {\bibinfo  {journal} {Appl. Phys. Lett.}\ }\textbf {\bibinfo
  {volume} {128}},\ \bibinfo {pages} {072106} (\bibinfo {year}
  {2026})}\BibitemShut {NoStop}%
\end{thebibliography}%

\end{document}